\documentclass[fleqn,usenatbib]{mnras}

\usepackage{xcolor}
\usepackage{newtxtext,newtxmath}
\usepackage[T1]{fontenc}

\DeclareRobustCommand{\VAN}[3]{#2}
\let\VANthebibliography\thebibliography
\def\thebibliography{\DeclareRobustCommand{\VAN}[3]{##3}\VANthebibliography}

\usepackage{graphicx}	% Including figure files
\usepackage[normalem]{ulem}
\usepackage{amsmath}	% Advanced maths commands
\usepackage{multirow}
\usepackage{subcaption}
\usepackage{comment}
\newcommand{\mbf}[1]{\mathbf{#1}}

\newcommand{\pc}[3]{${#1}^{+{#2}}_{-{#3}}$}

\newcommand{\cc}[1]{\left({#1}\right)}
\newcommand{\rr}[1]{\left[{#1}\right]}
\newcommand{\be}{\begin{equation}}
\newcommand{\ee}{\end{equation}}
\def\bear#1\ear{\begin{align}#1\end{align}}

\newcommand{\f}{\frac}

\newcommand{\e}{\mathrm{e}}

\renewcommand{\mathbf}[1]{\mbox{\boldmath $#1$}}

\newcommand{\snr}{S/N}
\newcommand{\spt}{SPT-3G}
\newcommand{\planck}{{\it Planck}}
\newcommand{\script}{\texttt{SCRIPT}}

\defcitealias{reichardt21}{R21}
\defcitealias{jain23}{J23}
\defcitealias{raghunathan23}{R23}

\title[Astrophysics with kSZ power spectrum]{Probing the Physics of Reionization Using kSZ Power Spectrum from Current and Upcoming CMB Surveys}

\author[Jain et al.]{
Divesh Jain$^{1}$\thanks{djain@ncra.tifr.res.in}, 
Tirthankar Roy Choudhury$^{1}$\thanks{tirth@ncra.tifr.res.in},
Srinivasan Raghunathan$^{2}$\thanks{srinirag@illinois.edu}
and
Suvodip Mukherjee$^{3}$\thanks{suvodip@tifr.res.in}
\\
$^{1}$ National Centre for Radio Astrophysics, Tata Institute of Fundamental Research, Pune 411007, India\\
$^{2}$ Center for AstroPhysical Surveys, National Center for Supercomputing Applications, Urbana, IL 61801, USA\\
$^{3}$ Department of Astronomy \& Astrophysics, Tata Institute of Fundamental Research, 1, Homi Bhabha Road, Colaba, Mumbai 400005, India\\
}
\date{\today}

\pubyear{2023}

\begin{document}
\label{firstpage}
\pagerange{\pageref{firstpage}--\pageref{lastpage}}
\maketitle

\begin{abstract}
The patchiness in the reionization process alters the statistics of Cosmic Microwave Background (CMB), with the kinematic Sunyaev-Zel’dovich (kSZ) effect in the CMB temperature power spectrum being a notable consequence. In this work, we aim to explore the potential of future kSZ power spectrum measurements in inferring the details of the reionization process. In this pursuit, we capitalize on the recent developments in foreground mitigation techniques using the Cross-Internal Linear Combination (Cross-ILC) technique, which enables robust detection of the kSZ power spectrum with signal-to-noise ($\snr$) roughly $20-30\sigma$ in this decade by \spt{} and Simons Observatory (SO); and $\ge80\sigma$ by CMB-S4 -- substantially improving on the recent evidence for kSZ binned at $\ell=3000$ using SPT-SZ+SPTpol surveys. We use a fiducial kSZ power spectrum along with realistic error bars expected from the above technique for SPT-3G, SO, and CMB-S4 to constrain the parameter space for a physical model of reionization. We find that with the improved error bars it will be possible to place stringent constraints on reionization using solely the Cross-ILC recovered SPT-3G kSZ without imposing any prior on $\tau$ in the Bayesian inference. Notably, high-fidelity kSZ measurements from CMB-S4 coupled with $\tau$ measurements through LiteBIRD will enable unprecedented constraint on the midpoint of reionization with an error bar of $\sim 0.25$ and the duration of  reionization with an error bar at $\sim 0.21$ exclusively using CMB data. This study highlights the need to capture kSZ power spectrum on a broad range of multipoles to gain insights into the inhomogeneous reionization era.
\end{abstract}


%%%%%%%%%%%%%%%%%%%%%%%%%%%%%%%%%%%%%%%%%%%%%%%%%%

%%%%%%%%%%%%%%%%% BODY OF PAPER %%%%%%%%%%%%%%%%%%

\section{Introduction}
Reionization is an important epoch in the evolution of the Universe, as the birth of the first stars started the end of the dark ages, initiating the ionization and heating of the hydrogen in the Intergalactic Medium (IGM). Understanding the process of reionization involves understanding the local physics of the formation and evolution of ionizing sources and the non-local physics of radiation transport \citep{2001PhR...349..125B,LoebFurlanetto+2013}. It started around the redshift of $z\sim 20-30$ and ended at $z\sim 5-6$ \citep{2006AJ....132..117F,2015MNRAS.447.3402B,2018MNRAS.479.1055B,2019MNRAS.485L..24K,2021MNRAS.501.5782C}, and represents the last phase transition of the IGM. Furthermore, simulations and analytical studies of the reionization era suggest that the process of reionization is patchy, indicating a spatially inhomogeneous ionization fraction during reionization \citep{2000ApJ...530....1M,2001PhR...349..125B,2004ApJ...613....1F,2005MNRAS.363.1031F,2007ApJ...671....1T,2007ApJ...654...12Z,2011MNRAS.411..955M,2018MNRAS.481.3821C,2021MNRAS.500..232P}. 

Cosmic Microwave Background (CMB) serves as the backlight from the past, capturing signatures about the ionization state of the Universe and its patchiness as they traverse the reionization era \citep{1998ApJ...508..435G,1999ASPC..181..227H,2000ApJ...529...12H,2016ASSL..423..227R,2022GReGr..54..102C}. The integrated evolution of the global ionization history of IGM during reionization has been captured through measurement of Thomson scattering optical depth $\tau$ by \cite{PlanckCollaboration2018}. While the patchiness in the ionization fraction, among other modulations, induces small-scale temperature anisotropy in the CMB referred to as the kinematic Sunyaev-Zeldovich (kSZ) signal, the evidence for which has been reported by \citet[hereafter \citetalias{reichardt21}]{reichardt21}; \cite{ gorce22}. In the future, we expect further high-fidelity observations of kSZ from future CMB experiments like the South Pole Telescope (\spt, \citealt{benson14, bender18}), Simons Observatory (SO, \citealt{so19}) and CMB-S4 \citep{cmbs419} while the most sensitive measurement of $\sigma(\tau) = 0.002$ (at cosmic variance limits) is expected from space-based telescopes LiteBIRD \citep{suzuki2018litebird} and PICO \citep{hanany2019pico} using the reionization bump in the CMB E-mode power spectrum.

The total kSZ encompasses two phases, namely the homogeneous and patchy kSZ phases. The contribution arising from the free electrons in the haloes of a fully ionized low redshift Universe is referred to as the homogeneous kSZ signal \citep{1986ApJ...306L..51O}. On the other hand, the contribution arising from the fluctuations in the ionization fraction during reionization is referred to as the patchy kSZ signal \citep{1998ApJ...508..435G,1999ASPC..181..227H}. Given the intricate nature of the reionization process, determining the strength of patchy kSZ is a difficult task. Broadly, this depends on the timing, duration, and morphology of the ionized bubbles during reionization and one needs to capture it from physics-driven reionization models \citep{Park_2013,2013ApJ...776...83B,2021MNRAS.500..232P,gorce22,2022ApJ...927..186T,2023ApJ...943..138C,2023MNRAS.526.3170N}. Detection of patchy kSZ, in its entirety, will enable crucial insights into the details of the patchy reionization era.
Nevertheless, the anticipated enhancement in the quality of kSZ data from upcoming high resolution CMB experiments \citep{benson14, so19, cmbs419}, necessitates a concerted effort on handling the systematic errors in these measurements. In particular,  the residual bias from the astrophysical foreground in these measurements is becoming increasingly important, as elaborated in \citet[hereafter \citetalias{raghunathan23}]{raghunathan23}, it should no longer be ignored.

In a recent development aimed at robustly extracting the kSZ signal, \citepalias{raghunathan23} presented a novel Cross-Internal Linear Combination (Cross-ILC) technique, using the cross-spectrum from two foreground-nulled maps, to minimize the residual bias from the astrophysical foregrounds, namely the cosmic infrared background (CIB) and the thermal Sunyaev-Zel’dovich  (tSZ) signals. Although this led to lower SNR than standard minimum variance ILC, the foreground systematics (CIB and tSZ) were significantly suppressed.
This resulted in robust bandpower estimates, as evidenced by bandpower errors across a range of multipoles for current and upcoming telescopes (see Fig. A1 of \citetalias{raghunathan23}). 
It was found that with the Cross-ILC approach, the total kSZ power spectrum can be measured at very high significance: $20-30\sigma$ in this decade by \spt{} and SO; and $\ge80\sigma$ by CMB-S4.

 In this work, we explore the potential of extracting kSZ signal from future experiments using the Cross-ILC technique to infer details on the reionization process. In \citet[hereafter \citetalias{jain23}]{jain23}, we developed a framework to compute reionization observables and CMB anisotropies self-consistently for a physical model of reionization. This allowed us to forecast constraints on the history as well as the patchiness in our reionization model when confronted with fiducial kSZ power (at central value $\ell=3000$, binned with $\Delta \ell=300$) from upcoming telescopes. However, with the potential to assess the shape of the power spectrum through the Cross-ILC technique in both current and future experiments, we can improve our understanding of the allowed reionization history as well as present stringent forecasts on the nature of ionizing sources. 
 
In order to simulate the patchy reionization kSZ we employ our self-consistent framework at the core of which is
a numerically efficient and explicitly photon-conserving semi-numerical model of reionization, Semi-numerical Code for ReIonization with PhoTon-conservation (\script{},\cite{2018MNRAS.481.3821C}). The
homogeneous kSZ contribution is derived using the cosmological scaling relations introduced in \cite{2012ApJ...756...15S} which is based on the biased matter power spectrum to describe the low-redshift spatial distribution of free
electrons. Following this, we simulate a mock kSZ data (allowed by current CMB estimates), with realistic error bars, as anticipated from the Cross-ILC technique for \spt, SO (Goal configuration), and CMB-S4 (wide survey from Chile), and confront it against kSZ predictions from our physical model, under a Bayesian framework, to gain insights into the patchy reionization era.

The paper is organized as follows: In Section \ref{sec:signalnsims}, we provide a brief description of the kSZ signal and discuss the simulation of reionization used to evaluate this signal in our model. In Section \ref{sec:motivation}, we motivate the need to capture the kSZ power spectrum over a range of multipoles to gain insights into the source properties as well as the reionization history. In Section \ref{sec:crossilc}, we discuss the Cross-ILC already defined above technique for extracting the kSZ signal, emphasizing its comparative advantages, over other prevalent methods such as the Template-based and standard Minimum Variance (MV-ILC) techniques. In Section \ref{sec:compare_const}, we confront our model with existing and upcoming probes of $\tau$ and kSZ signal and obtain constraints on the reionization model, with particular emphasis on the improvements in forecasts offered by CMB experiments as a result of kSZ extracted from the Cross-ILC technique.  In Section \ref{sec:discussandconclude}, we discuss and summarize our conclusions.

It should be noted that throughout this study, we have fixed the cosmological parameters to $[\Omega_m, \Omega_b, h, n_s, \sigma_8] = [0.308, 0.0482, 0.678, 0.961, 0.829]$ \citep{PlanckCollaboration2014} which is consistent with \cite{planck20_cosmo}.

\section{Simulating kSZ signal in the CMB}\label{sec:signalnsims}

Patchiness in the reionization process leads to a line of sight $\hat{n}$ dependence on the Thomson scattering optical depth 
\begin{equation}
    \tau(\hat{n},
    \chi)=\sigma_T \bar{n}_{H}  \int^\chi_0 \frac{d\chi^\prime}{{a^\prime}^2} x_e(\hat{n},\chi^\prime),
\end{equation}
where the free electron fraction is defined as $x_e\equiv n_e/n_H$. Here, $\bar{n}_H$ is the mean comoving number density of hydrogen, and $\sigma_T$ is the Thomson scattering cross-section. The optical depth to the last scattering $\tau(\hat{n})=\tau(\hat{n},\chi_{\rm LSS})$ surface can be evaluated by integrating the above integral till $\chi_{\rm LSS}$ corresponding to the redshift of last scattering surface given by $z_{\rm LSS}$.

This patchiness ($x_e(\hat{n},\chi)$)
introduces patchy reionization imprints on CMB. Specifically, when CMB photons Thomson scatter off ionized bubbles (local ionization fluctuations $\Delta x_e(\hat{n},\chi)$) with a net bulk velocity $\mathbf{v}(\hat{n},\chi)$, a secondary temperature anisotropy on the CMB is imprinted as a result of the Doppler shifting of the photons called the patchy kinematic Sunyaev Zeldovich (kSZ) signal. The temperature anisotropy induced along the line of sight hence depends on the dimensionless ionized momentum field as 
\bear
\frac{\Delta T}{T_0}=-\sigma_T\bar{n}_H\int d\chi  (1+z)^2 e^{-2\tau(\chi)} \mathbf{q}(\hat{n},\chi)\cdot \hat{n},
\ear
where, the dimensionless ionized momentum field $\mathbf{q}$ is defined as
\begin{equation}
    \mathbf{q}=x_e(1+\delta)\frac{\mathbf{v}}{c}.
\end{equation}

Under Limber's approximation, the angular power spectrum of this patchy secondary temperature anisotropy, the patchy kSZ signal, is given as \citep{PhysRevLett.88.211301,Park_2013,2021MNRAS.500..232P}. \\

\bear
C_\ell^{\mathrm{kSZ,reion}} &= \left(\sigma_T \bar{n}_{H} T_0\right)^2 \int\f{ d\chi \e^{-2 \tau(\chi)}}{2a^4\chi^2} ~ P_{q_\perp}(k = (l+1/2)/\chi,\chi).
\label{eq:ksz_lim}
\ear
In this equation, $P_{q_\perp}$ represents the power spectrum of the transverse component of the Fourier transform of the momentum field $\mbf{q}(\mbf{k},z)$ defined as $\mbf{q}_\perp(\mbf{k}, z) = \mbf{q}(\mbf{k}, z) - (\mbf{q}(\mbf{k}, z) \cdot \mbf{k}) \mbf{k}/k^2$.

\subsection{Simulating the kSZ signal with SCRIPT}\label{sec:scriptksz}

In this work, we use a semi-numerical scheme of reionization that is explicitly photon-conserving called
Semi-numerical Code for ReIonization with PhoTon-conservation, abbreviated as \script \citep{2018MNRAS.481.3821C}, to simulate the ionization maps during reionization and consequently the CMB observables of reionization, $\tau$, and patchy kSZ.  In addition to its time efficiency, \script{} generates power spectrum for a fluctuating field that is convergent at large scales across map resolutions. This feature differentiates \script{} from other semi-numerical models of reionization based on the popular excursion set approach \citep{2007ApJ...669..663M,2011MNRAS.411..955M,2008MNRAS.386.1683G}, preventing any inference bias when opting to work with coarser map resolutions.

The first step is to generate dark matter snapshots at these redshifts. For a fixed set of cosmological parameters, we generate dark matter snapshots at $\Delta z=0.1$ for redshifts $5 \leq z \leq 20$ employing the 2LPT prescription in MUSIC \citep{hahn2011multi} for box length of $512 ~ h^{-1}$ Mpc with $512^3$ particles. The collapsed mass fraction in haloes is computed using a subgrid prescription based on the conditional ellipsoidal collapse model \citep{2002MNRAS.329...61S}, see \cite{2018MNRAS.481.3821C} for more details of the method.

 In order to generate an ionization map at a redshift, \script{} relies on the two input parameters describing the property of ionizing sources. These are $M_{\rm min}$,  representing the minimum mass of haloes that can host ionizing sources, and $\zeta$ which signifies the effective ionizing efficiency of these sources. Using these parameters \script{} generates a map of ionized hydrogen fraction $x_{\mathrm{HII}}(\mbf{x},z)$. For this study, our parameter of interest is the free electron fraction

\be
x_e(\mbf{x},z) = \chi_{\mathrm{He}}~x_{\mathrm{HII}}(\mbf{x},z)~\Delta(\mbf{x},z),
\ee
where, $\chi_{\mathrm{He}}$ is the correction factor to account for free electrons from ionized Helium and $\Delta(\mbf{x},z)$ corresponds to the dark matter overdensity. In our analysis, we consider $\chi_{\mathrm{He}}=1.08$ for $z>3$ corresponding to contribution from singly-ionized Helium and $\chi_{\mathrm{He}}=1.16$ for $z<3$ to account for free electron contribution from doubly ionized Helium. To enable us to capture the small-scale inhomogeneities, ionization maps using \script{} are generated with the best possible resolution of $2 ~ h^{-1}$ Mpc. 

Modeling the reionization and hence the emerging CMB anisotropies, is contingent on the parameterization we assume for $M_{\rm min}(z)$ and $\zeta(z)$ across redshift. Because of a lack of knowledge about how ionizing sources evolve in the reionization era, we assume an intuitive redshift-based power-law model for $M_{\rm min}$ and $\zeta$. The parameterization is thus considered as the following

\begin{table}
    \def\arraystretch{1.6}
    \centering
    \caption{Specifications of priors for the free parameters used during Bayesian inference of reionization parameters}
    \begin{tabular}{|c|c|c|}
    \hline
    Parameter & range & nature \\
    \hline
    $\log_{10} \zeta_{\rm 0}$     & (0, $\infty$) & uniform\\
    $\log_{10} M_{\rm min,0}$  & [7.0, 11.0] & uniform \\
    $\alpha_\zeta$ & (-$\infty$, $\infty$) & uniform\\
    $\alpha_M$ & (-$\infty$, 0] & uniform\\
    \hline
    \end{tabular}
    \label{tab:paramprior}
\end{table}

\begin{equation}
    \zeta(z) = \zeta_0 \left(\f{1 + z}{9}\right)^{\alpha_{\zeta}};~~ M_{\mathrm{min}}(z) = M_{\mathrm{min},0} \left(\f{1 + z}{9}\right)^{\alpha_{M}}.
\end{equation}
Here, $M_{\mathrm{min},0}$ is the minimum mass of haloes that can contribute to the ionizing process at redshift $z=8$ while $\zeta_0$ is the ionizing efficiency of these sources at $z=8$. The parameters $\alpha_M$ and $\alpha_\zeta$ correspond to indices of the power law. Therefore, the reionization process can be completely described by the four free parameters $\mathbf{\theta}\equiv[\log_{10} \zeta_0,\log_{10} M_{\mathrm{min},0},\alpha_\zeta,\alpha_M]$, using which one can compute ionization maps with \script{}  at redshifts of interest. In Table \ref{tab:paramprior}, the priors for these free parameters, as used during Bayesian inference, are presented. We exclude models exhibiting unphysical ionizing efficiency, specifically those where $\zeta(z)> 10^5$ and $\zeta(z)< 10^{-1}$. Additionally, in our analysis, we only consider
reionization histories which are complete by redshift ($z\geq5$), consistent
with constraints presented in \citep{2011MNRAS.415.3237M,2019MNRAS.485L..24K,2020MNRAS.499..550Q}. The evaluation of the patchy kSZ signal considered in this work requires an evaluation of the perpendicular component of the momentum field power spectrum from the ionization maps. When simulating kSZ power for finite box sizes, one misses out on
the velocity field contribution of wavemodes with wavelengths larger than the size
of the box in $P_{q_\perp}$, consequently underestimating the patchy kSZ signal. In this work, we correct for this missing large-scale wave mode contribution using the analytical calculation of the missing power in $P_{q_\perp}$
presented in \cite{Park_2013}. For details on the specific implementation, we encourage the reader to refer to Appendix A in \citetalias{jain23}. 

However, the total kSZ signal, in addition to the patchy kSZ, receives a contribution from the motion of ionized halos in the post-reionization period called the homogeneous kSZ. In this work, we account for the homogeneous kSZ from the cosmological scaling relations presented in \cite{2012ApJ...756...15S} for their cooling and star formation (CSF) model (refer to Table 3 in \cite{2012ApJ...756...15S}). These scaling relations depend on cosmological parameters $H_0$, $\sigma_8$ and $\Omega_b$ as well as astrophysical parameters $\tau$ and redshift corresponding to the end of reionization $z_{\rm rei}$. When we explore the parameter space during Bayesian inference, as cosmological parameters are fixed in this study, we compute the homogeneous kSZ ($D_\ell^{\mathrm{kSZ, post-reion}}$), using  $\tau$ and $z_{\rm rei}$ evaluated for the corresponding reionization model. Therefore for each sample of our free parameter $\mathbf{\theta}$, we self-consistently evaluate the total kSZ power spectrum  as $D_\ell^{\mathrm{kSZ}}=D_\ell^{\mathrm{kSZ, reion}}+ D_\ell^{\mathrm{kSZ, post-reion}}$.

\begin{figure}
	\includegraphics[width=\columnwidth,trim={0 0.5cm 0 0cm},clip]{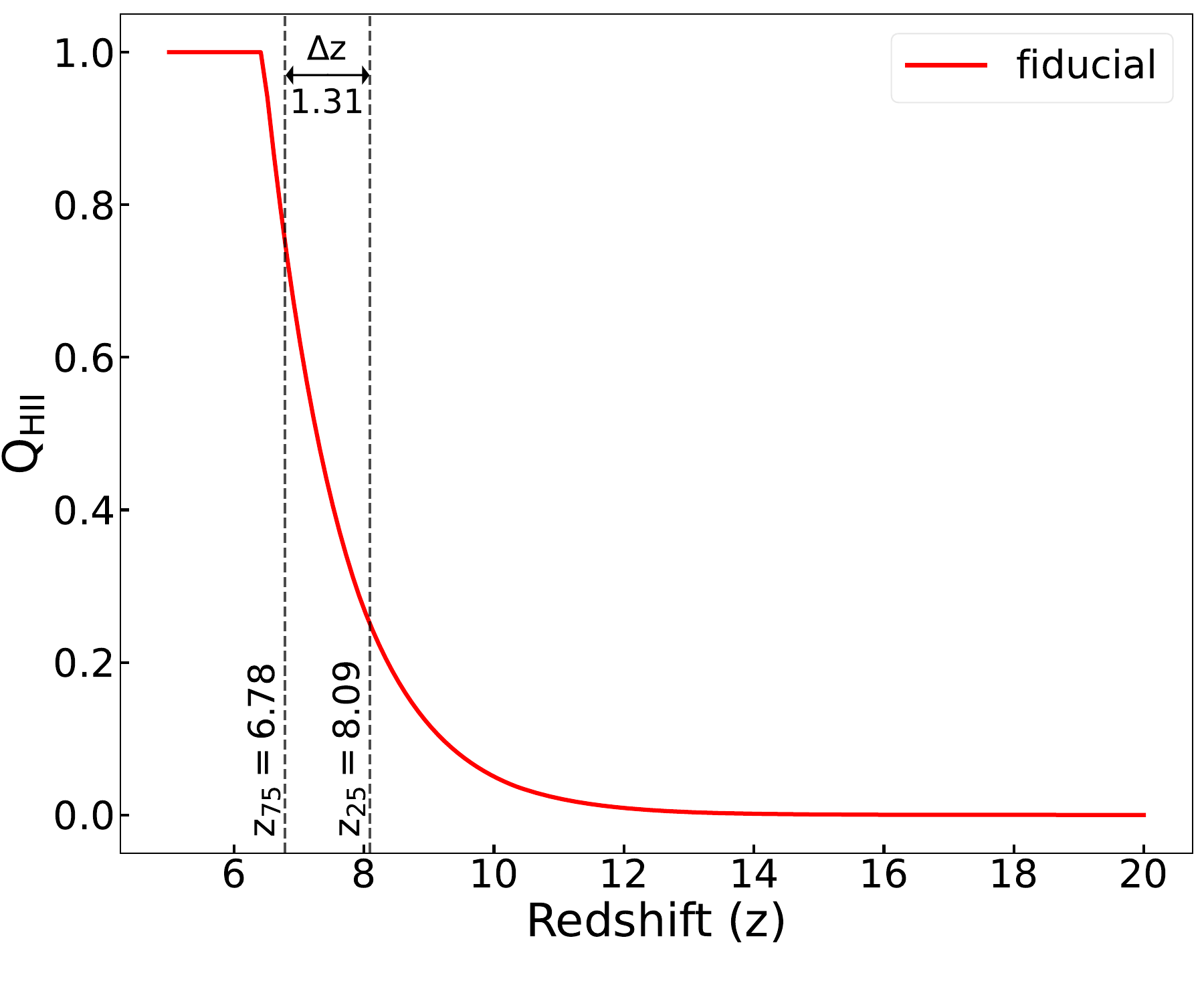}
    \caption{Redshift evolution of mass-averaged ionized fraction $Q_{\rm{HII}}(z)$ for fiducial model of reionization. The dashed lines denote $z_{25}$ and $z_{75}$, the 25\% and 75\% reionization redshifts at which $Q_\text{HII}$ equals 0.25 and 0.75 respectively. $\Delta z = z_{25} - z_{75} = 1.31$ corresponds to the duration of our fiducial reionization model.}
    \label{fig:ionizationfrac}
\end{figure}

For fiducial mock data, we consider the best-fit model of reionization as the fiducial model of reionization, obtained corresponding to the Bayesian inference carried out on the above model of reionization in \citetalias{jain23} using recent CMB measurements. The inference used constraints on  $\tau=0.054$ with $\sigma^{\mathrm{obs}}_\tau=0.007$ \citep{planck20_cosmo} and the kSZ signal  \citepalias{reichardt21} at $D^{\mathrm{kSZ,obs}}_{\ell=3000}=\ell(\ell+1)C^{\mathrm{kSZ,obs}}_{\ell=3000} =3 ~\mu K^2$ with a $\sigma^{\mathrm{kSZ,obs}}_{\ell=3000}=1~\mu K^2$. The best-fit model, parameterization was obtained to be $\rr{\log_{10} M_{\mathrm{min},0}=9.73,\log_{10} \zeta_0=1.58, \alpha_M=-2.06, \alpha_\zeta=-2.01}$. The value of $\tau$ for this fiducial model is $0.054$ and kSZ signal amplitude is $D^{\rm kSZ}_{\ell=3000}=3 \mu K^2$.  The redshift evolution of the global mass-averaged ionization fraction $Q_{\rm{HII}}(z) \equiv \langle x_{\mathrm{HII}}(\mbf{x},z)~\Delta(\mbf{x},z) \rangle$ for this model is shown in Figure \ref{fig:ionizationfrac}.

\section{Motivation}\label{sec:motivation}

\begin{figure}
    \centering
	\subfloat[Variation in reionization contribution to kSZ with minimum mass of halo at $z=8$, $\log M_{\rm min,0}$.]{\includegraphics[width=\columnwidth,trim={0 1.3cm 0 0.6cm},clip]{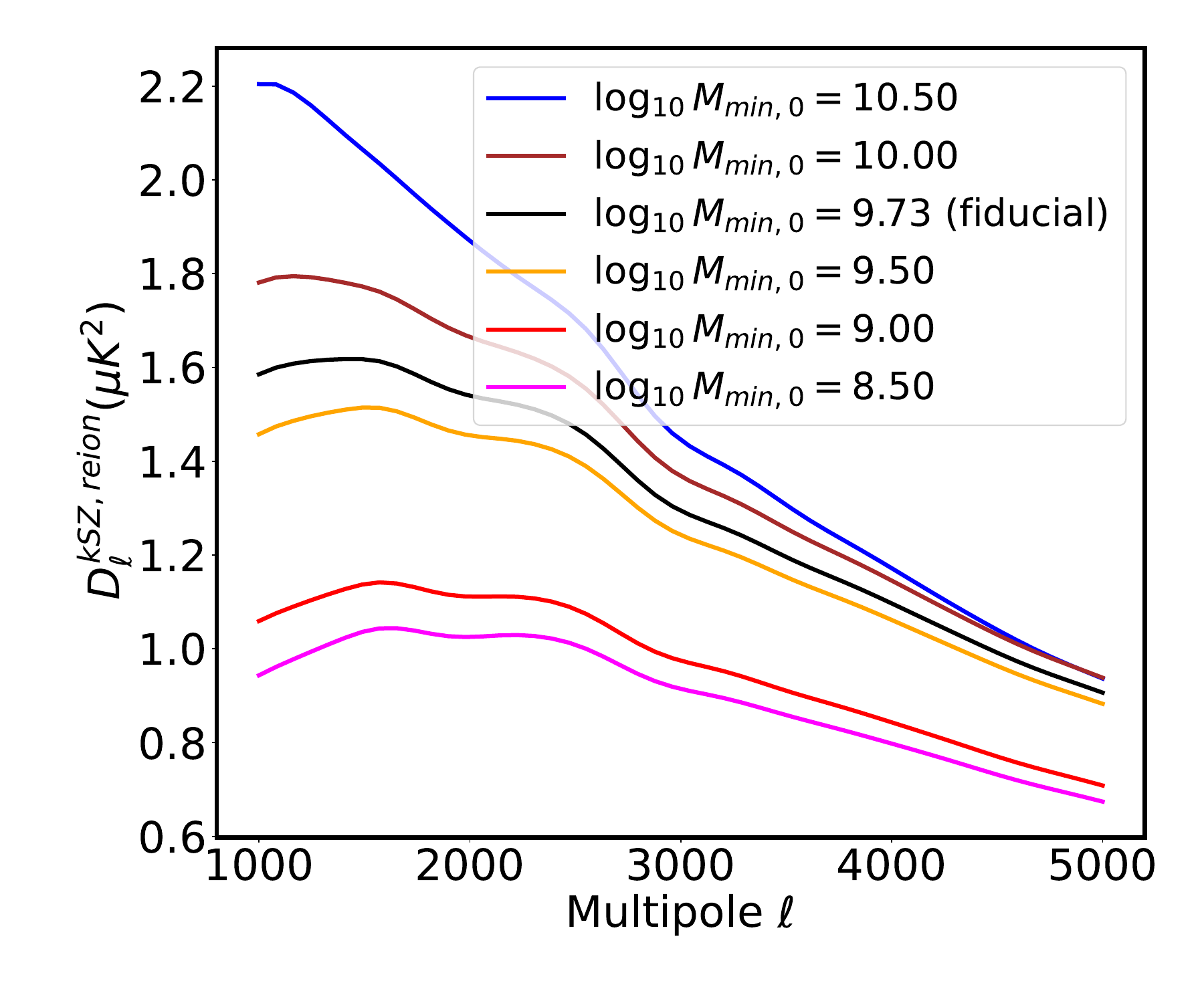}}\label{fig:varksz_patchy_mass}
	\subfloat[Variation in reionization contribution to kSZ with midpoint of reionization, $\Delta z$.]{\includegraphics[width=\columnwidth,trim={0 1.3cm 0 0.6cm},clip]{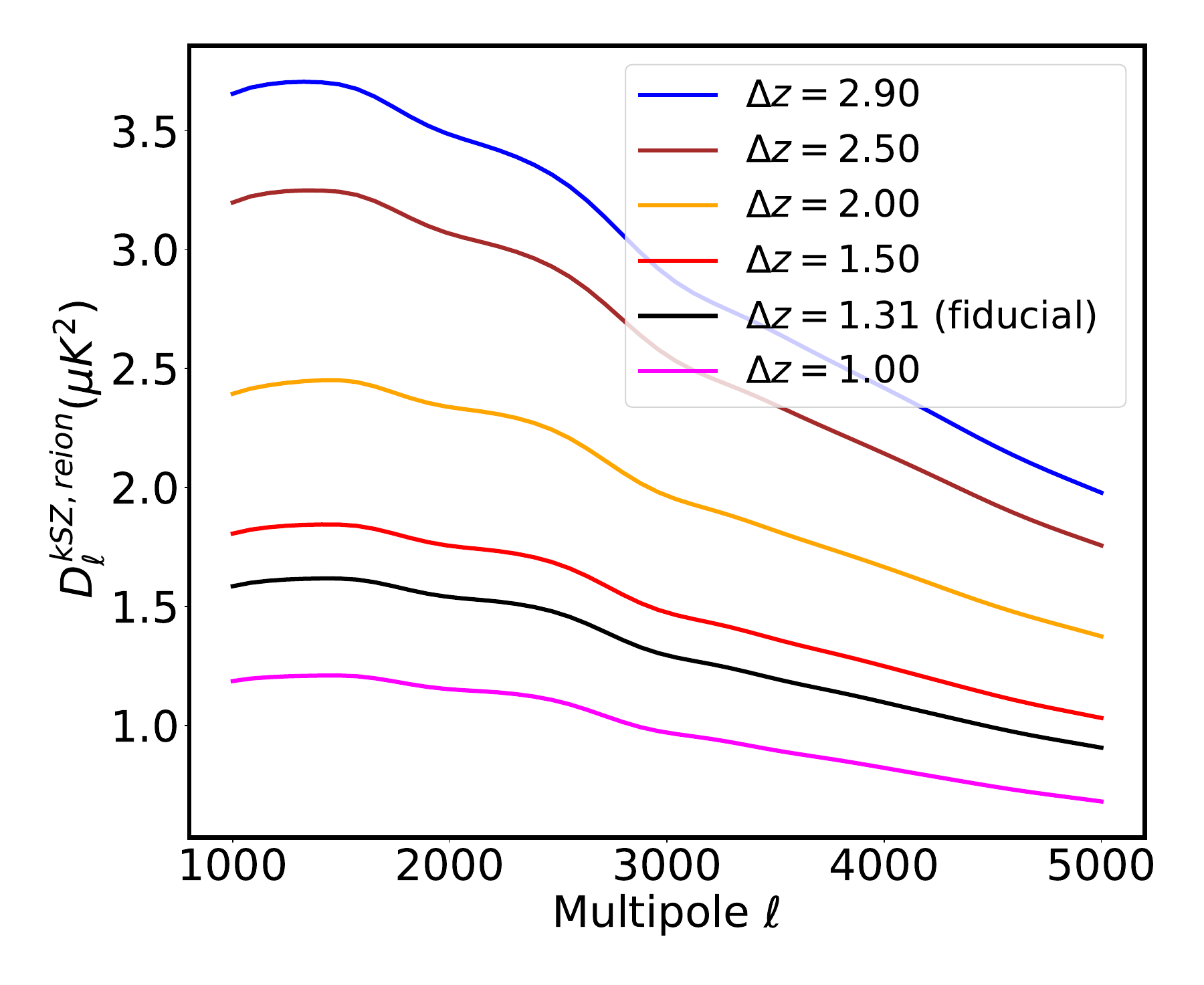}}\label{varksz_patchy_deltaz}
 \subfloat[Variation in reionization contribution to kSZ with midpoint of reionization, $z_{50}$.]{\includegraphics[width=\columnwidth,trim={0 1.3cm 0 0.6cm},clip]{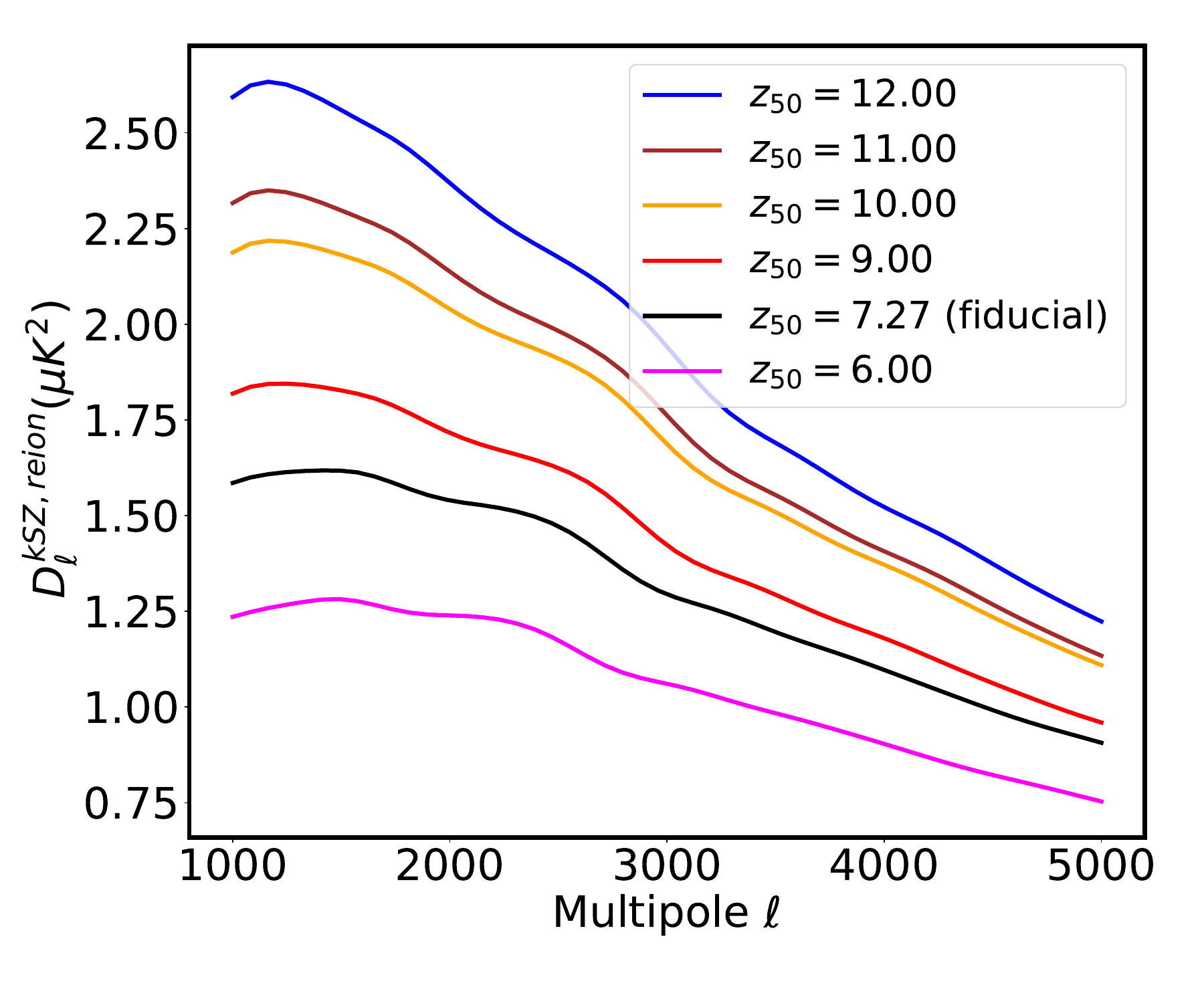}}\label{varksz_patchy_z50}

    \caption{Variation in patchy kSZ power spectrum for different reionization model parameters, the minimum mass of halos hosting ionizing sources $\log_{10}M_{\rm min}$, the duration of reionization $\Delta_z$ and the mid-point of reionization $z_{50}$. Each panel alters one of three parameters: $\log_{10} M_{\rm min,0}$, $\Delta z$, or $z_{50}$, deviating from the fiducial model while maintaining the others at similar values. The fiducial parameter values are  [$\log_{10} M_{\rm min,0}=9.73$ (representing the minimum mass of halos at redshift $z=8$ with a power-law index of $\alpha_M=-2.06$ denoted as $\alpha_M$), $\Delta z=1.31$, and $z_{50}=7.27$].}
    \label{fig:varksz_patchy}
\end{figure}

The shape and amplitude of the kSZ power spectrum modulate in response to physical conditions during reionization. Contrary to this, previous works \citep{2005ApJ...630..643M,2012ApJ...756...65Z,2013ApJ...776...83B} have suggested
that midpoint of reionization, $z_{50}$, and the duration of reionization, $\Delta z$ to be sufficient to evaluate the patchy kSZ contribution from reionization.  However, as demonstrated in  \cite{2020A&A...640A..90G,2021MNRAS.500..232P}, such simplifications are not only inadequate but may yield misleading estimations of kSZ. Additionally, the shape of the power spectrum will typically capture the sizes of ionized regions which is missed if we limit ourselves to the amplitude at $\ell = 3000$. In this regard, we discuss how details of reionization impact the kSZ and why is it necessary to explore the amplitude of the signal beyond $\ell=3000$.

Fundamentally, two attributes characterize the kSZ power spectrum: its shape, and its amplitude. The shape is primarily determined by the size of the ionized regions "bubbles" around the sources of ionizing photons (primarily galaxies) while the amplitude largely depends on the inhomogeneity in IGM, the duration of reionization, and the timing of reionization \citep{2012ApJ...756...65Z,2020A&A...640A..90G,2021MNRAS.500..232P}.

Let's discuss the shape of the kSZ signal. The size of the ionized region, assuming no prior history of ionization, is determined by the number of ionizing photons deposited into the IGM by the source within the region. This number depends on the ionizing efficiency of the source which scales with the mass of haloes that hosts this source. The scattering of CMB photons at these ionized regions would correspond to the multipole (at which power roughly peaks) being inversely proportional to the size of the ionized region. Consequently, the multipole, at which power peaks, correlates with the mass of the halo. Therefore, if reionization is dominated by sources with higher mass haloes, the resulting kSZ signal will likely peak at larger scales i.e. smaller multipoles. 

The amplitude of the signal generally increases with the inhomogeneity, the duration of 
reionization and the timing of reionization. Considering an identical reionization history reionization history, a scenario driven by larger halo masses will cause patchier bubble distribution leading to a higher kSZ signal \citep{2020A&A...640A..90G,2021MNRAS.500..232P}. In the context of our reionization framework, this is controlled using a parameter $\log_{10} M_{\rm min}$ which determines the minimum mass of the halo that can host a reionizing source at any given redshift. The top panel of Figure \ref{fig:varksz_patchy} describes the evolution in the reionization contribution as a function of the minimum mass of halo $\log_{10} M_{\rm min}$ at redshift 8 denoted through $\log_{10} M_{\rm min,0}$, predicted by our model. The black curve signifies the fiducial model for this study, referring to the best fit to current constraints from \planck's $\tau$ and \spt's kSZ measurements. For this model, the duration of reionization is $\Delta z=1.31$ while the midpoint occurs at $z_{50}=7.27$.  Notably, all the curves in the top panel even with varying $\log_{10} M_{\rm min,0}$, correspond to similar duration and midpoint of reionization as the best-fit model, resulting in a similar optical depth to reionization at $\tau\approx 0.054$. To focus on the effect of the minimum mass of halo on reionization kSZ, we consider a power-law evolution of $\alpha_M=-2.06$ across all models, consistent with the fiducial one. We find that aligning with our expectation the increase in the minimum mass of halo results in a patchier reionization scenario and increases the kSZ contribution. Further, as  $\log _{10} M_{\rm min,0}$ increases, the peak of the angular power spectrum shifts to larger scales or smaller multipole $\ell$ indicating a comparative increase in bubble sizes as larger mass haloes dominate the reionization scenario.

Considering the effect of the duration of reionization, the power in the kSZ signal increases with duration as the number of interactions between CMB photons and the ionized regions tends to grow with the duration. In the middle panel of Figure \ref{fig:varksz_patchy}, we see that when keeping the minimum halo mass and the mid-point of reionization preserved at $\log_{10}M_{\rm min,0}=9.73; \alpha_M=-2.06$ and mid-point of duration at $z_{50}=7.27$, the increase in the duration of reionization increases the reionization contribution of kSZ. A similar increase is also reflected if the average redshift of reionization is higher, depicted through the bottom panel in Figure \ref{fig:varksz_patchy}. Under such circumstances, the mean density of the Universe is comparatively higher and the same duration leads to an increased kSZ power \citep{2005ApJ...630..643M,2012ApJ...756...65Z}, as reflected in the bottom panel. Note that as the midpoint of reionization varies, even as the $\log_{10}M_{\rm min}$ and $\Delta z$ are similar for all the models, the optical depth to reionization in such a case would deviate from $\tau=0.054$.

In the above discussion, we have not invoked the role of the velocity distribution of ionized regions. The velocity would determine the positive enhancements or negative cancellations of temperature anisotropy along the line of sight. We also emphasized the variation in the patchy reionization contribution to kSZ as reionization models vary. However, in the upcoming sections, we will self-consistently incorporate the post-reionization kSZ to evaluate the total kSZ contribution. Ascertaining kSZ across a range of multipoles will enable us to probe the growth of structure and, importantly, the details of the reionization process. In this work, this is consistently facilitated by the numerical model of reionization which allows us to reliably evaluate the kSZ signal shape, even in extreme scenarios.

\section{Mitigation of foreground contamination using  Cross-ILC technique}\label{sec:crossilc}

Even though kSZ is a powerful probe of late-time astrophysics, capturing the shape of kSZ is difficult and is largely dependent on handling foreground systematics. To handle the foregrounds, the standard approach has been to use simulation-based templates and jointly fit the templates with the goal of mitigating the undesired astrophysical component (\citetalias{reichardt21}, \citealt{gorce22}). However, if these templates are misestimated, it can bias the kSZ estimation.

An alternative approach is to use the standard Internal Linear Combination approach, which aims to minimize the total variance in the observed data $M_{\ell}$. This is achieved by applying frequency-channel dependent weights $w_\ell$ as represented by:

\begin{equation}
    S_\ell = \sum^{N_{\rm ch}}_{i=1}w^i_\ell M^i_\ell.
\end{equation}
Here, $N_{\rm ch}$ represents the frequency channels, and $w^i_\ell$ are the weights associated with each frequency channel $i$ and multipole $\ell$. If the matrix $ \mathbf{C}_{\ell} $, of dimensions $ \rm N_{\rm ch} \times \rm N_{\rm ch} $, encapsulates the covariance between maps across frequencies for a given multipole $ \ell $, the weights are tuned to yield a minimum-variance (MV) signal map by minimizing $w^\dagger_\ell \mathbf{C}_{\ell} w_\ell$. \citetalias{raghunathan23} demonstrate that in the context of kSZ extraction, the MV ILC technique introduces significant bias due to the inaccuracies in modeling tSZ and CIB signals.

To improve the signal extraction, a modified approach— the constrained ILC (cILC) \citep{2011MNRAS.410.2481R} offers a more robust way to generate bias-free maps. Unlike the standard ILC, cILC employs weight functions that simultaneously minimize variance and null-specific frequency responses. The idea is to minimize the variance of $w^\dagger_\ell  \mathbf{C}_{\ell} w_\ell$ such that some undesired frequency responses can be nullified. Let us denote the frequency response vector $ \mathcal{F} $ as a combination of the desired signal $ A_{S} $ to be extracted and the undesired components that are to be nullified, given as $ [A_{S}\ B_{S}\ C_{S}\ ...\ Z_{S}] $ and let N denote the vector that dictates which components in $\mathcal{F}$ are to be nullified (usually represented by zeroes) and which are to be retained (usually represented by ones). Mathematically, we would like to find the weights $w_\ell$ which satisfies

\begin{equation}
    \min_{w_{\rm \ell}} \left( w_{\rm \ell}^{\dagger} \mathbf{C}_{\rm \ell} w_{\rm \ell} \right) , \ \ \ \ \text{s.t.}\ \ \ \ \mathcal{F}^{\dagger} w_{\rm \ell} = N.
\end{equation}
This can be solved using Lagrangian multipliers to achieve

\begin{equation}
w_{\rm \ell}^{\rm cILC} = {\mathbf{C}}^{-1}_{\rm \ell} \mathcal{F} \left(\mathcal{F}^{\dagger} {\mathbf{C}_{\rm \ell}}^{-1} \mathcal{F} \right)^{-1} N.
\end{equation}

Extending this approach to minimize the total residual arising from CIB and tSZ, \citetalias{raghunathan23} presented the Cross-ILC with the aim to robustly detect the kinetic Sunyaev-Zeldovich (kSZ) power spectrum. In this approach, two distinct cILC maps are created: the first nullifies the tSZ and the second nullifies the CIB. Subsequently, the cross-spectrum of these two maps is computed for analysis. This adaptation minimizes the total residuals arising from CIB and tSZ taking into account the ${\rm CIB} \times {\rm tSZ}$ cross-correlations. While this method has a noise penalty and thus a lower S/N than the MV ILC, it produces more robust estimates of the kSZ  in a direction closer to its unbiased measurements. In this work, we exploit the development in kSZ extraction with Cross-ILC to forecast the insights we anticipate on reionization models with future kSZ experiments.

\section{Parameter Forecasts}\label{sec:compare_const}
In this section, we forecast constraints on reionization parameters using the fiducial kSZ power spectrum with Cross-ILC error bars. We additionally provide forecasts for the joint datasets of fiducial $\tau$ and fiducial kSZ. The forecasts obtained will represent the best possible constraints on ionizing source parameters and reionization history enabled by the combination of polarization and temperature measurements for current and upcoming telescopes.  
\subsection{Forecasts with fiducial kSZ power spectrum extracted using Cross-ILC technique for SPT-3G}\label{sec:compare_const_spt}

\begin{figure*}
    \centering
    \includegraphics[width=\linewidth]{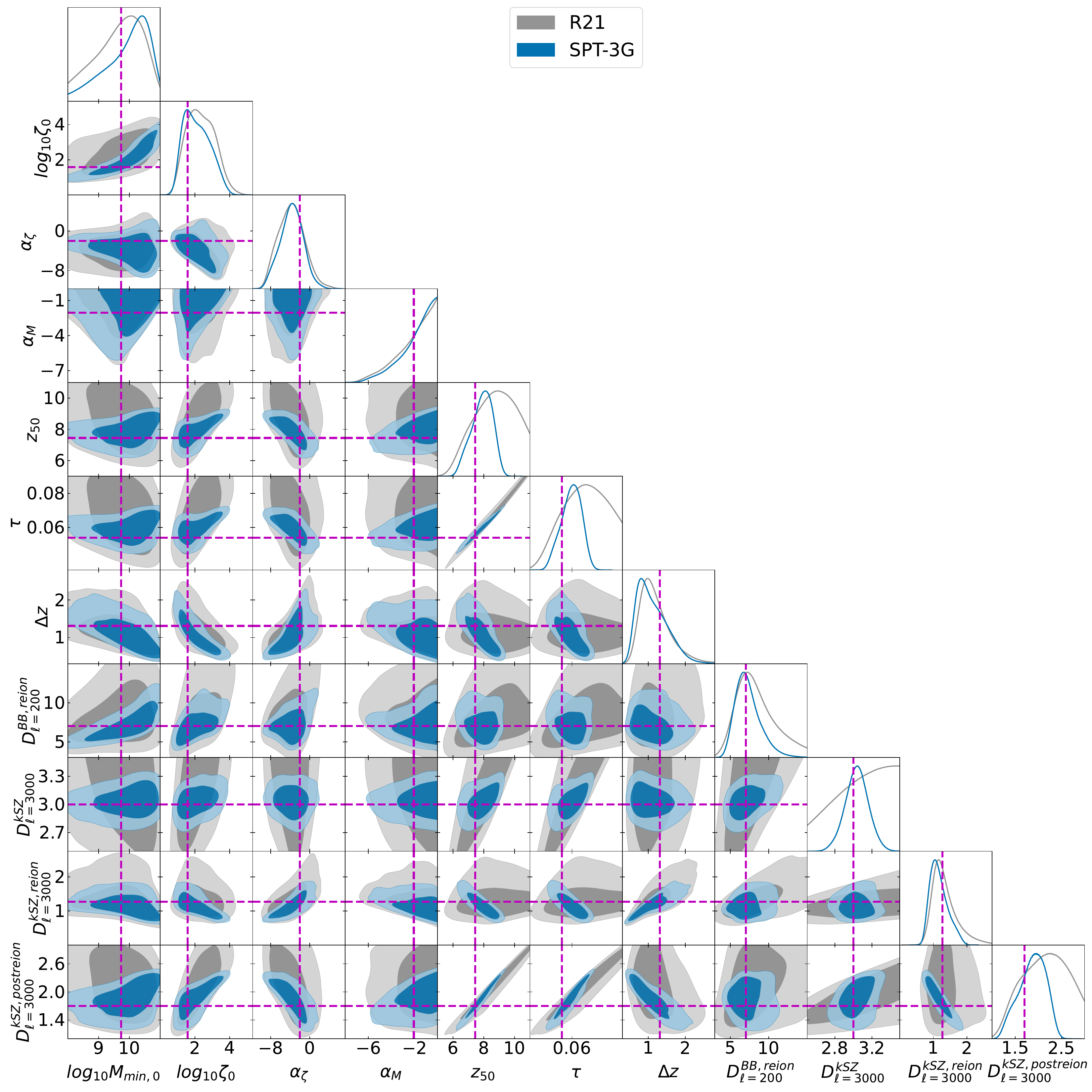}
    \caption{Comparison of the 2D posterior distribution of reionization model parameters obtained from MCMC analysis using current observations from R21 and forecasts for SPT-3G. The dashed lines denote the input model used to generate the fiducial data for forecasting.}
    \label{fig:alljustkszfig3a}
\end{figure*}

\begin{figure*}
    \centering
    \includegraphics[width=\linewidth]{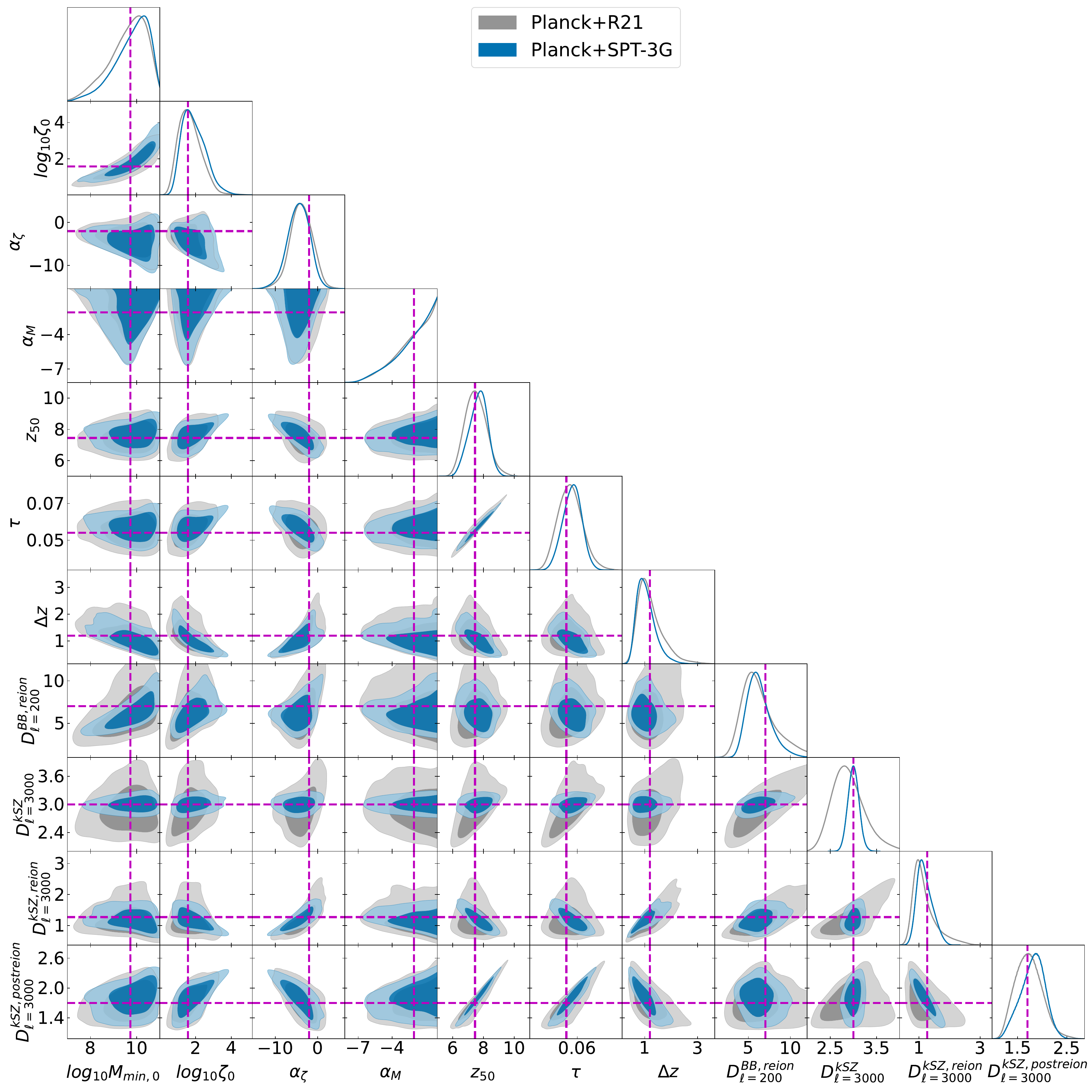}
    \caption{Comparison of the 2D posterior distribution of reionization model parameters obtained from MCMC analysis using current observations for the case Planck+R21
and forecasts for case Planck+SPT-3G. The dashed lines denote the input model used to generate the fiducial data for forecasting.}
    \label{fig:alltaukszfig3b}
\end{figure*}

\setlength{\tabcolsep}{1.2pt}
\begin{table}
    \fontsize{7.3}{10}\selectfont
    %\centering
    \caption{Comparison of forecasts ($68\% $ limits) for reionization model parameters obtained from MCMC analysis using improved Cross-ILC error bars on fiducial SPT-3G power spectrum through the case SPT-3G and Planck+SPT-3G have been presented. The first four rows correspond to the free parameters of the model while the rest of the parameters are the derived parameters. The second column refers to the model of reionization used to generate the fiducial data to forecast constraints for the SPT-3G and Planck+SPT-3G case. For comparison, the constraints obtained for our reionization model from the \planck{} $\tau$ and  SPT  kSZ data (\citetalias{reichardt21}) have been presented through cases R21 and Planck+R21.}
    %|p{2.2cm}|p{0.6cm}|p{2.4cm}|p{2.6cm}|
    \begin{tabular}{|c|c|c|c|c|c|}
    \hline
    Data  &  & R21 &SPT-3G & Planck+R21 &Planck+SPT-3G \\
    Parameter  & Input&  & Forecast   &  & Forecast    \\
    \hline
    $\log_{10} \zeta_0$& 1.58 & $2.16^{+0.78}_{-0.90}$ & $2.09^{+0.53}_{-0.95}$ &\pc{1.70}{0.49}{0.76} & \pc{1.90}{0.49}{0.77}\\[0.075cm]
    $\log_{10} M_{\mathrm{min,0}}$ & 9.73 &\pc{9.44}{1.09}{0.48} & $9.82^{+0.99}_{-0.35}$ & \pc{9.65}{1.02}{0.49} & \pc{9.79}{0.95}{0.41}\\[0.1cm]
    $\alpha_\zeta$ & -2.01 & \pc{-4.42}{2.65}{2.83}  & $-3.87^{+2.52}_{-2.37}$ & \pc{-3.81}{2.58}{2.52} &\pc{-4.46}{2.72}{2.41} \\[0.1cm]
    $\alpha_M$ & -2.06  & $>-2.63$ &$>-2.33$ & $>-2.78$  & $>-2.60$   \\[0.1cm]
    \hline
    Derived  &&\\
    Parameters &&\\
    $\tau$ & 0.054 & \pc{0.0721}{0.0146}{0.0195}  & $0.0596^{+0.0070}_{-0.0054}$ & \pc{0.0559}{0.0062}{0.0067} &\pc{0.0571}{0.0055}{0.0048}\\[0.1cm]
    $z_{\mathrm{50}}$& 7.27 & \pc{9.01}{1.61}{1.68} &\pc{7.89}{0.85}{0.60}&\pc{7.49}{0.69}{0.68} &\pc{7.65}{0.64}{0.53} \\[0.1cm]
    $\Delta z$    & 1.31 & \pc{1.20}{0.21}{0.50} &$1.10^{+0.23}_{-0.49}$& \pc{1.19}{0.27}{0.53} &\pc{1.10}{0.22}{0.46}  \\[0.1cm]
    $D^{\mathrm{kSZ}}_{l=3000}$($\mu K^2$) & 3.00 & \pc{3.58}{0.58}{0.76}&$3.04^{+0.14}_{-0.13}$& \pc{2.90}{0.26}{0.41} & \pc{3.00}{0.12}{0.12}\\[0.1cm]
    $D^{\mathrm{kSZ,reion}}_{l=3000}$($\mu K^2$) & 1.33 & \pc{1.34}{0.20}{0.44}&$1.17^{+0.17}_{-0.33}$& \pc{1.20}{0.13}{0.44} & \pc{1.19}{0.19}{0.30}\\[0.1cm]
    $D^{\mathrm{kSZ,postreion}}_{l=3000}$($\mu K^2$) & 1.67 & \pc{2.23}{0.56}{0.60}&$1.87^{+0.36}_{-0.23}$& \pc{1.71}{0.24}{0.30} & \pc{1.81}{0.27}{0.22}\\[0.1cm]
    $D^{BB,\mathrm{reion}}_{\ell=200}$(${\rm n K^2}$)   & 7.03 & \pc{8.91}{1.33}{3.82} & $7.17^{+1.07}_{-2.14}$ &  \pc{6.60}{1.13}{2.73} & \pc{6.32}{0.97}{1.69} \\[0.1cm]
    \hline
    \end{tabular}
    \label{tab:forecast-SPT3GCROSS_ilc_COMPARISON}
\end{table}

The most sensitive measurement of optical depth to reionization $\tau$ was presented in \cite{PlanckCollaboration2018} at \mbox{$\tau=0.054\pm 0.007$} 
based on the full  mission TT ($2 \leq \ell \leq 2500$), TE ($30 \leq \ell \leq 2000$), and notably, the low-multipole E-mode bump in EE ($2 \leq \ell \leq 2000$) data. The data also included the \planck{} CMB lensing signal ($8 \leq \ell \leq 400$). The first $3\sigma$ measurement for the kSZ signal was presented in \citetalias{reichardt21} and is at $D^{\mathrm{kSZ,obs}}_{\ell=3000}=\ell(\ell+1\ )C^{\mathrm{kSZ,obs}}_{\ell=3000} =3 ~\mu K^2$ with a $\sigma^{\mathrm{kSZ,obs}}_{\ell=3000}=1~\mu K^2$. To extract this kSZ signal, the temperature signal from the 2500 $\mathrm{deg}^2$ SPT-SZ and 500 $\mathrm{deg}^2$ SPT-pol surveys in the range $2000 \leq \ell  \leq 11,000$ (corresponding to angular scales of $1' \lesssim \theta \lesssim 5'$) were employed. While \citetalias{reichardt21} provided a measurement of kSZ power at $\ell=3000$, the Cross-ILC technique on available SPT-3G dataset will enable access to power spectrum bins between $\ell \in [2500,5000]$, as elaborated in this section. To emphasize the role of Cross-ILC error bars we will begin by presenting constraints on the available CMB data through the case R21 (model constraints obtained from current kSZ data) and the case Planck+R21 (model constraints obtained from the combination of current $\tau$ and kSZ data). Following this we will delve into a thorough comparison with forecasted constraints through cases SPT-3G and Planck+SPT-3G where ``SPT-3G" refers to fiducial Cross-ILC extraction from available SPT-3G dataset.

We obtain parameter constraints under the Bayesian framework by employing the MCMC sampler in the \texttt{Cobaya} framework \citep{Torrado_2021} to sample the free parameters of our reionization model $\mathbf{\theta}\equiv\rr{\log_{10} \zeta_0,\log_{10} M_{\mathrm{min},0},\alpha_\zeta,\alpha_M}$. Each set of sample $\mathbf{\theta}$ yields the derived parameters $\rr{\tau,D^{\rm kSZ}_{\ell}}$. We compare these derived parameters accordingly to available and fiducial data sets for each of the above cases and obtain the posteriors on the reionization parameters. The likelihood used for the analysis of Planck+R21 and Planck+SPT-3G, considering the appropriate choice of $\ell_{bins}$ and observed and mock data choice, takes the following form:

    \begin{equation}
    -2\log \mathcal{L}  =  \cc{\frac{\tau-\tau^{\mathrm{obs}}}{\sigma^{obs}_\tau}}^2 +  \sum_{\ell_{bins}}\cc{\frac{D^{\rm kSZ}_{\ell}-D^{\rm kSZ,obs}_{\ell}}{\Sigma^{\rm kSZ}_{\ell}}}^2. 
    \end{equation}
\begin{figure*}
	\subfloat[Variation with minimum mass of halo at $z=8$]{\includegraphics[width=0.337\textwidth, trim={1.3cm 1cm 1.2cm 3.5cm}, clip]{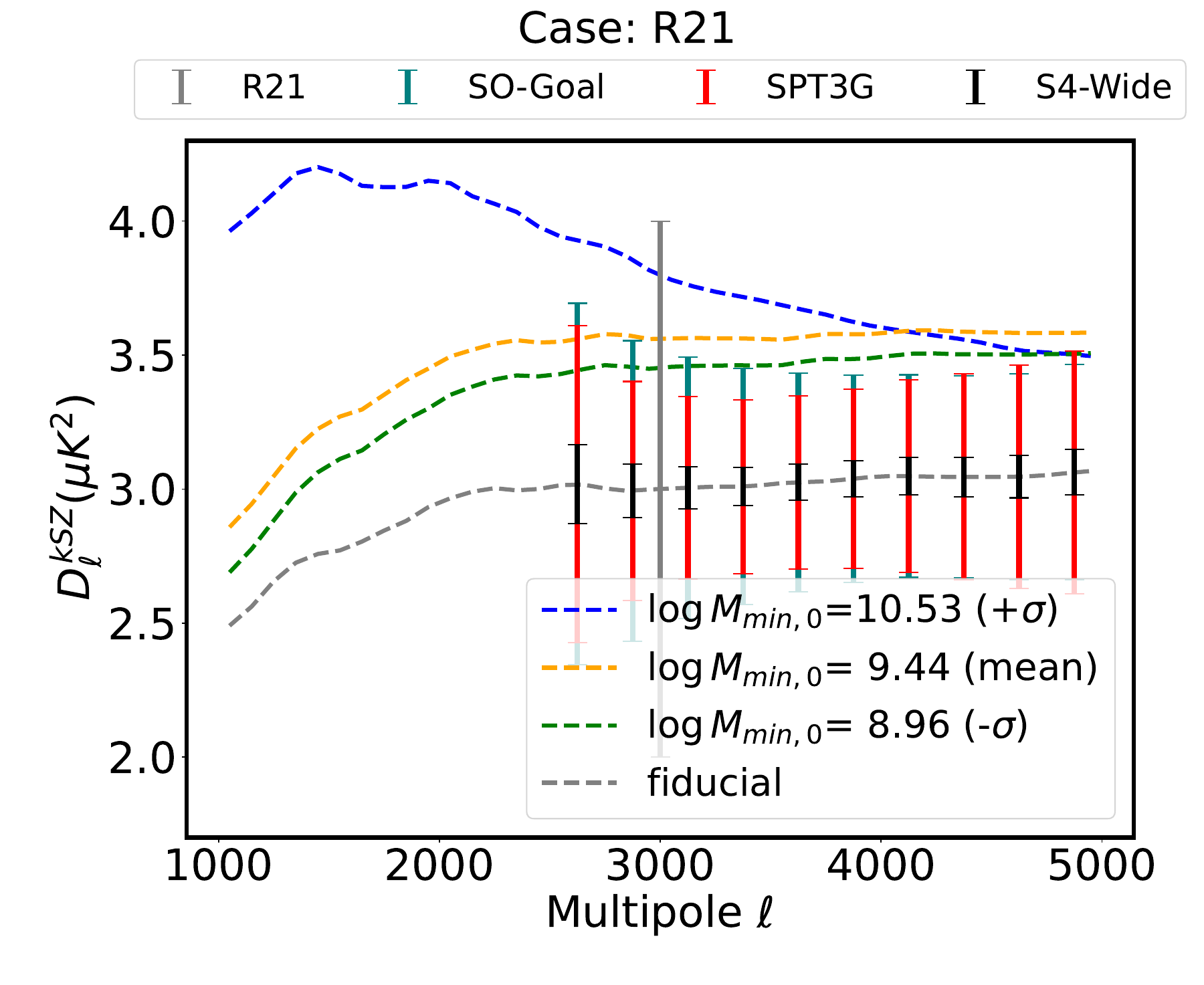}}\label{jkszvarM}
	\subfloat[Variation with duration of reionization.]{\includegraphics[width=0.327\textwidth,  trim={1.3cm 1cm 0.9cm 0.2cm}, clip]{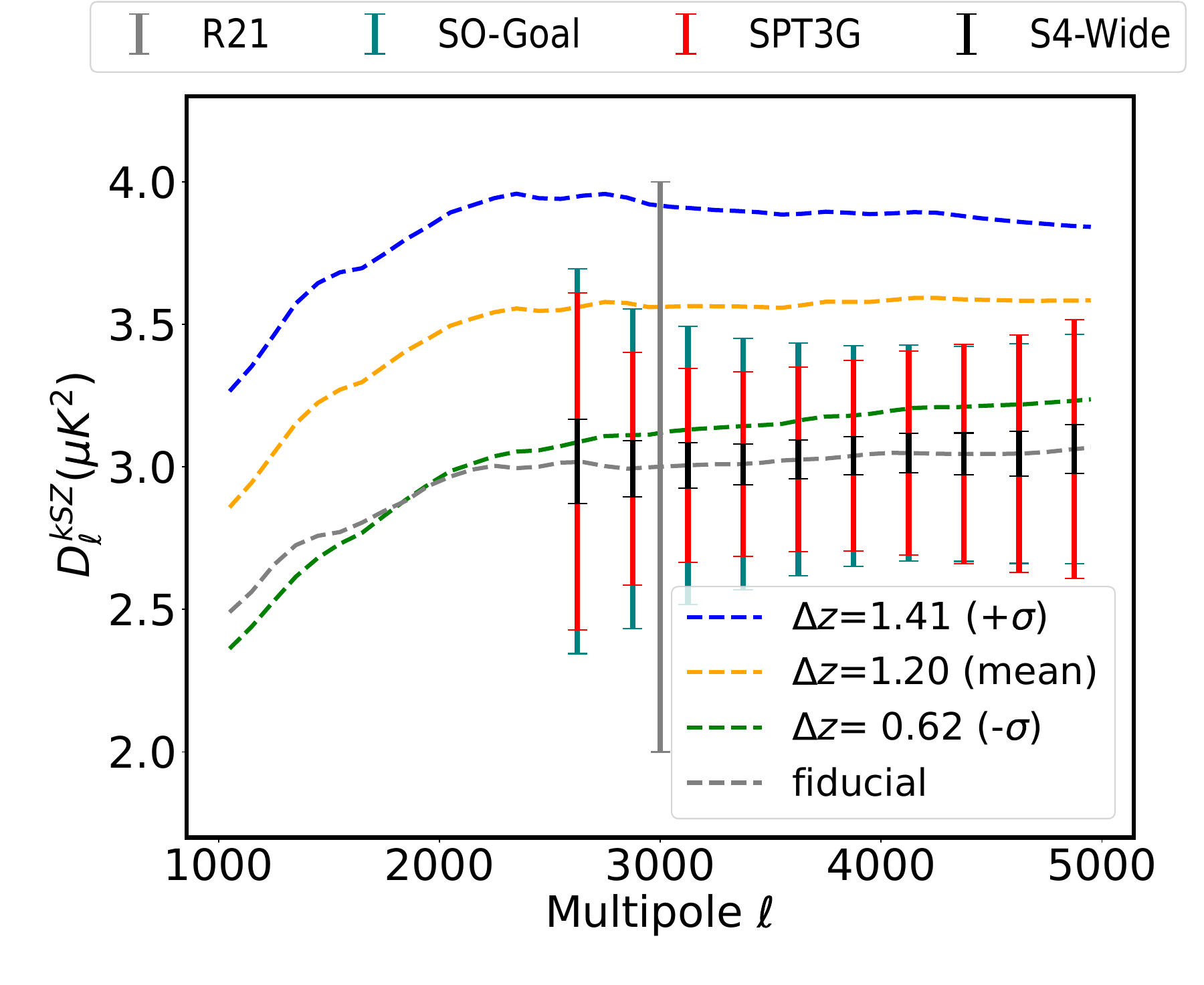}}\label{jkszvardeltaz}
 	\subfloat[Variation with mid-point of reionization.]{\includegraphics[width=0.327\textwidth, trim={1.3cm 1cm 1.2cm 1.9cm}, clip]{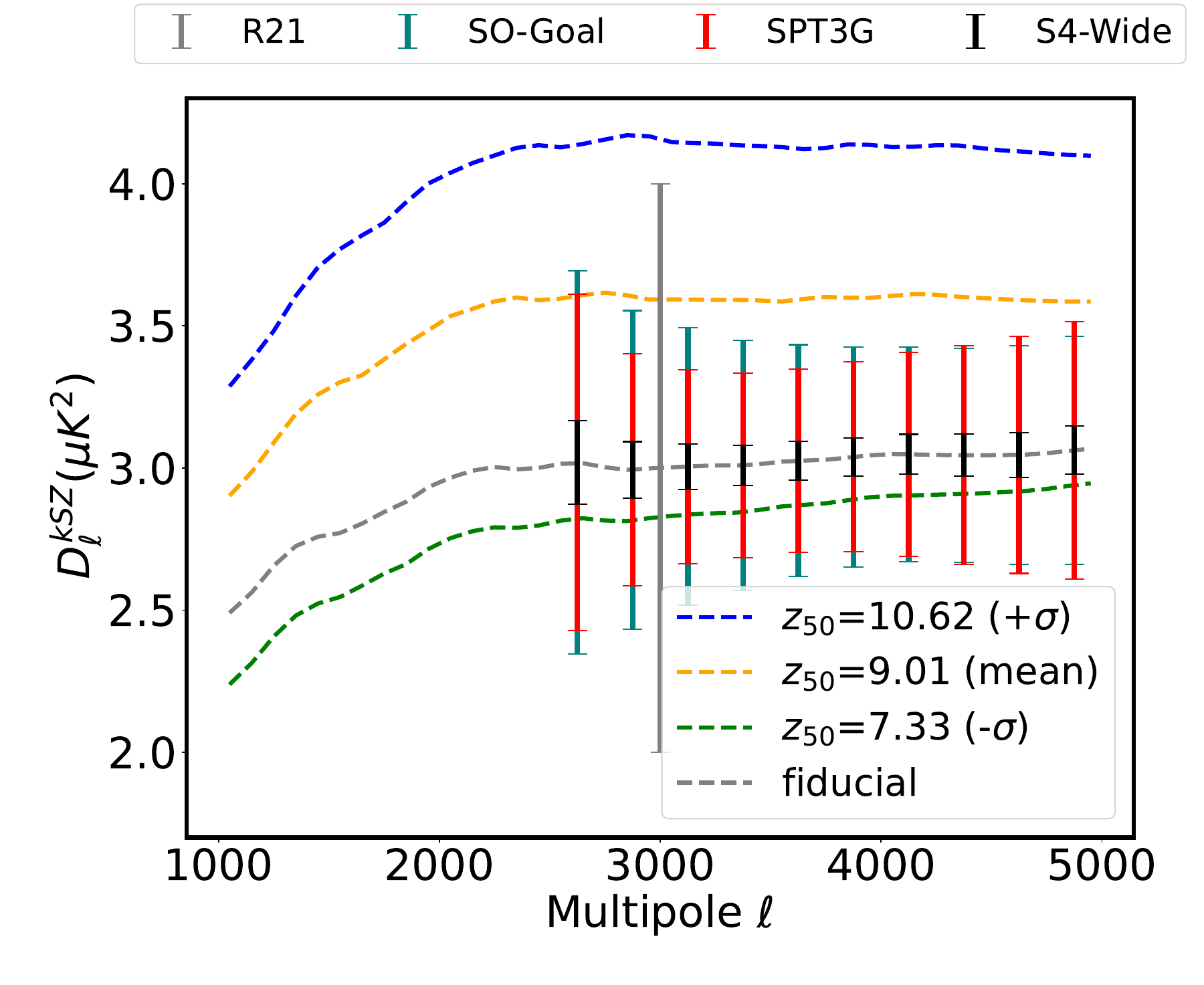}}\label{jkszvarz50}
    \caption{kSZ power variation with the reionization models, presented through figure legends (mean, $+\sigma$, $-\sigma$), allowed by \citetalias{reichardt21} kSZ measurement (presented through grey error bar). Each panel alters one of three parameters: $M_{\rm min,0}$, $\Delta z$, or $z_{50}$ by $+\sigma/-\sigma$, deviating from the mean model while maintaining the others at similar values permissible by the chains. Additionally, the fiducial model (best-fit, \citetalias{jain23}) allowed by the current CMB estimates from \planck{}'s $\tau$ and \citetalias{reichardt21}'s kSZ measurement is presented. The band power error bars from the Cross-ILC \citepalias{raghunathan23} technique are presented for SO-Goal, SPT3G, and S4-Wide CMB experiment overlaid on the fiducial model.}
    \label{fig:jvarkSZ}
\end{figure*}

For the case R21 and SPT-3G cases, we consider only the contribution from the kSZ-$\chi^2$ term appropriately in the likelihood. For all forecast-related cases, we introduce Gaussian random noise to the fiducial data set. This step is taken to simulate random errors originating from telescope noise, which our simulation cannot capture. We also provide forecasts without the addition of Gaussian Random Noise to the fiducial data. These results are available in Table \ref{tab:forecast-CROSS_ilc_COMPARISON_no_noise} in the Appendix for reference.

In addition to forecasts on the free parameters of the model, we forecast constraints on parameters associated with reionization history i.e. midpoint of reionization $z_{50}$ which corresponds to reionization redshift corresponding to mass-averaged ionization fraction of $Q_{\rm HII}=0.50$ and $\Delta z=z_{\rm 25}-z_{\rm 75}$, the width of reionization. Here, $z_{25}$ and $z_{75}$ refer to the reionization redshift which corresponds to mass-averaged ionization fraction of $Q_{\rm HII}=[0.25,0.75]$ respectively. We further present forecasts for CMB probes of reionization namely $\tau$, kSZ power (both patchy $D^{\rm{kSZ,reion}}_{\ell}$ and homogeneous $D^{\rm{kSZ,postreion}}_{\ell}$) at $\ell=3000$ and patchy $B$-mode power at $\ell=200$, $D^{BB ,{\rm reion}}_{\ell=200}$.

For each prescribed case, we present a comparison of the parameter constraints in Table \ref{tab:forecast-SPT3GCROSS_ilc_COMPARISON}. The two-dimensional posterior distribution for the case R21 is presented in  Figure \ref{fig:alljustkszfig3a} and for the case, Planck+R21 have been presented in Figure \ref{fig:alltaukszfig3b}. For the case R21 we find that we can place a constraint on $\log_{10} M_{\rm min,0}$ at $9.44^{+1.09}_{-0.48}$ and an $\alpha_\zeta$ constraint at $-4.42^{+2.65}_{-2.83}$. Notably, consistent with our previous findings in \citepalias{jain23} (also presented in Table \ref{tab:forecast-SPT3GCROSS_ilc_COMPARISON} for the case Planck+R21), \citetalias{reichardt21} data independently suggests $M_{\rm min} \gtrsim 10^9 M_\odot$  at redshift 8 with $68\%$ confidence, indicating suppression of star formation in lower mass haloes as a result of radiative feedback. $\alpha_\zeta$ prefers a negative constraint, which could be indicative of improvement in cooling and efficiency of star formation
or increased escape fraction at lower redshifts. 

\begin{figure}
    \centering
  \begin{subfigure}{\linewidth}
    \centering
    \includegraphics[width=\linewidth,trim={0cm 10.5cm 0 0cm},clip]{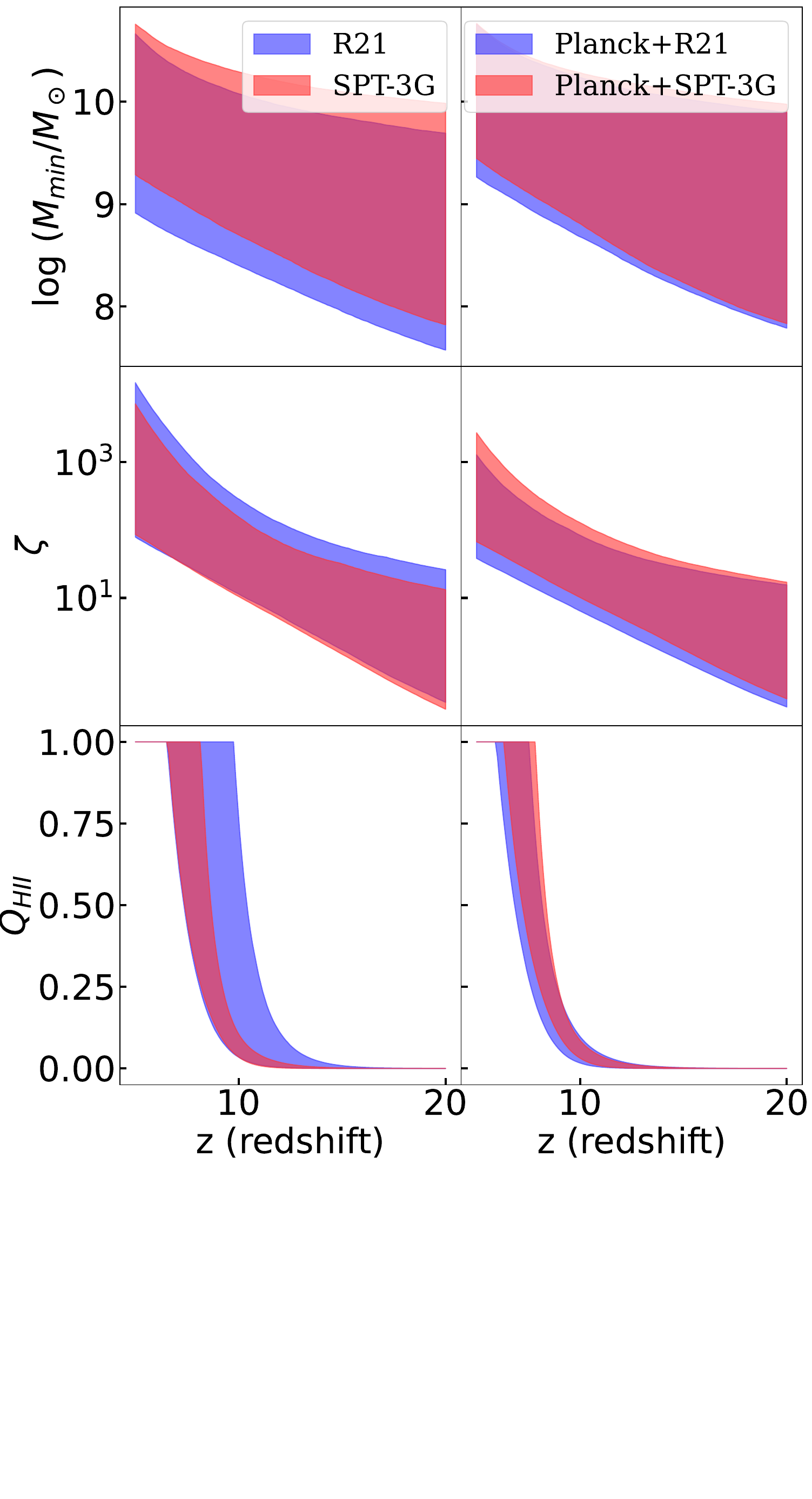}
    \label{fig:68pcqhii}
  \end{subfigure}
  \caption{Evolution of minimum mass of haloes hosting sources which contribute to reionization, the ionizing efficiency of the sources $\zeta(z)$ and the ionized mass fraction $Q_{\rm HII}$ for models in MCMC chains has been shown.
The left column presents results corresponding to the cases R21 and SPT-3G, while the right column shows the same quantities for the cases Planck+R21 and Planck+SPT-3G.}
  \label{fig:68pc}
\end{figure}

We find that our analysis with the SPT data \citepalias{reichardt21} imposes strong constraints on the duration of reionization, with $\Delta z=1.20^{+0.21}_{-0.50}$ at 68\% C.L. and $\Delta z = 1.20^{+1.57}_{-0.75}$ at 99\% C.L. These are consistent yet more stringent than the constraint on the duration of reionization from the analysis presented in \citetalias{reichardt21}. These tight constraints are attributable to the self-consistent evaluation of the reionization and post-reionization kSZ power within a physical model of reionization. However, within this model, constraints on ionizing efficiency are weak, permitting models with high ionizing efficiency, as reionization progresses. This leads to a preference for early reionization scenarios and consequently leads to a higher inference of $\tau={0.0721}^{+0.0146}_{-0.0195}$ and $z_{\rm 50}={9.01}^{+1.61}_{-1.68}$.

Further,$D^{\rm kSZ}_{\ell=3000}$ for our model is constrained at \pc{3.58}{0.58}{0.76} $\mu K^2$ which is consistent with \citetalias{reichardt21} data at $3 \pm 1$ $\mu K^2$.The reionization contribution to kSZ $D^{\rm kSZ, reion}_{\ell=3000}$ is constrained at $1.34^{+0.20}_{-0.44} \; {\rm \mu K^2}$ with the post-reionization contribution to kSZ is constrained at $D^{\rm kSZ, postreion}_{\ell=3000} = 2.23^{+0.56}_{-0.60} \; {\rm \mu K^2}$. It is important to note that the correlation observed between $z_{\rm 50}$ and $\tau$ with $D^{\rm kSZ, postreion}_{\ell=3000}$ in Figure \ref{fig:alljustkszfig3a}, although intuitive i.e. an earlier end of reionization entails a higher kSZ contribution from the post-reionization era, is a direct consequence of the cosmological scaling relations used to estimate the homogeneous kSZ \citep{2012ApJ...756...15S}. But, the large error bars on \citetalias{reichardt21} data do not allow for obvious signatures of correlations between $z_{50}$ and $\tau$ with the patchy kSZ amplitude, $D^{\rm kSZ, reion}_{\ell=3000}$, at multipole of $\ell=3000$ for the case of R21. However, there exists a clear correlation between $D^{\rm kSZ, reion}_{\ell=3000}$ and the duration of reionization $\Delta z$, clearly indicating that an increase in the duration of reionization leads to an increased patchy reionization contribution to kSZ.

Combining measurements of optical depth from \planck{}, referred as case Planck+R21 (also studied in \citetalias{jain23}), results in improved constraints of the ionizing source parameters.   Consequently, this enhancement in parameter constraints leads to more stringent error bars for the reionization history-associated parameters. The error bars on $z_{50}$ and $\Delta z$ are constrained at 0.69 and 0.40. Tighter constraints on ionizing source parameters further reflect as tighter constraints on $D^{\rm kSZ, postreion}_{\ell=3000}$ with an error bar of $\sim 0.27 {\rm \mu K^2}$ and $D^{\rm kSZ, reion}_{\ell=3000}$ with an error bar of $\sim 0.29 {\rm \mu K^2}$.

In Figure \ref{fig:jvarkSZ}, the variation of the constrained kSZ signal for case R21 has been presented. We show variation in the kSZ signal with the minimum mass of halo at redshift 8, $\log_{10} M_{\rm min,0}$, the duration of reionization, $\Delta z$, and the mean redshift of reionization, $z_{\rm 50}$ from the mean model parameters. It is important to note that for the mean (also standard deviation $-\sigma$) model across all the parameters, the kSZ power at $\ell=3000$, although consistent with \citetalias{reichardt21}, exceeds kSZ's mean measurement at 3$\mu K^2$ (2 $\mu K^2$). 
This arises as a result of the exclusion of models associated with reionization histories with very late and rapid completion of reionization, which inherently exhibits lower kSZ power, through priors on the end of reionization and unphysical ionizing efficiencies. The models for which $\zeta(z)<0.1$, transition from nearly zero ionizing efficiencies to a very rapid rise of ionizing photon deposition into IGM within an extremely narrow duration of reionization. These models are inconsistent with our physical understanding where we expect a smoother evolution of galaxy mass function over the Hubble time scale and the evolution of the escape of ionizing photons into the  IGM. Consequently, for our model, this limits preference for reionization models with a late and narrow duration of reionization, and hence their rejection results in the inference of a marginally higher mean kSZ signal.

Detection of kSZ across a range of multipoles will enable us to probe the growth of structure and, importantly, the details of the reionization process. kSZ extraction techniques like Cross-ILC will enable unbiased constraints on the kSZ signal at various multipole bins. We present the band power errors from the Cross-ILC technique \citepalias{raghunathan23} over the kSZ signal corresponding to our fiducial model of reionization in the Figure \ref{fig:jvarkSZ} for the ongoing experiment SPT-3G, as well as future kSZ observing experiments like the Simons Observatory and CMB-S4. With the improvement in sensitivity of kSZ measurement, it is unsurprising that measurement on kSZ over a range of multipoles will allow us to rule out reionization models that produce kSZ inconsistent with the error bars. This, in turn, will enable us to make stringent constraints on the ionizing source parameters as well as the reionization history. We leverage this argument to present forecasts on our physical model of reionization with fiducial data sets from the above experiments through the cases SPT-3G and Planck+SPT-3G. 

The two-dimensional posterior distribution for the case SPT-3G  is presented in  Figure \ref{fig:alljustkszfig3a} and for the case, Planck+SPT-3G have been presented in Figure \ref{fig:alltaukszfig3b}. Comparing the cases R21 and SPT-3G, we find that with information from the shape of the spectrum in the case SPT-3G,  tight error bars on our model's free parameters are obtained including constraints on the $\log_{10} \zeta_0$ and $\log_{10} M_{\rm min,0}$.  In the left column in Figure \ref{fig:68pc}, we present $68\%$ spread of the minimum mass of haloes $M_{\rm min}$ that can host ionizing sources, the ionizing efficiency of the sources $\zeta$ and the evolution of mass average ionization fraction from the MCMC chains. With SPT-3G forecasts, a significant part of parameter space for models corresponding to high ionizing efficiency and low minimum mass haloes is ruled out indicating a preference for late reionization scenarios (refer to bottom-left panel in Figure \ref{fig:68pc}). This additionally allows the possibility of constraining $\tau$ with error bars at $\sim 0.0062$ compared to the error bar obtained with R21 at $\sim 0.0171$. Notably, the error bars on $\tau$ from SPT-3G are comparable to the error bars of the best available measurement of $\tau$ at $0.054\pm 0.007$ through Planck. Consequently, this enables the possibility of establishing independent constraints on $\tau$ from small-scale CMB temperature measurements complementing the current inference derived from large-scale CMB polarization measurements. Stringent constraints on reionization source parameters and consequently the reionization history parameters results in stringent constraints on the kSZ amplitude, $D^{\rm kSZ}_{\ell=3000}$ at an error bar of $\sim 0.13 {\rm \mu K^2}$ while the individual contributions of patchy $D^{\rm kSZ, reion}_{\ell=3000}$ and homogeneous kSZ $D^{\rm kSZ, postreion}_{\ell=3000}$ are constrained with an error bar of $\sim 0.25 {\rm \mu K^2}$ and $\sim 0.30 {\rm \mu K^2}$ respectively. Additionally, preference for late and gradual reionization scenario as indicated bottom-left panel in Figure \ref{fig:68pc} implies a lower post-reionization contribution to kSZ and results in the expected anti-correlation between reionization contribution and post-reionization contribution to kSZ in Figure \ref{fig:alljustkszfig3a} to unravel. The constraints on patchy $B$-mode, $D^{ BB,{\rm reion}}_{\ell=200}$ improve from an error bar $\sim 2.57 {\rm nK^2}$ in the R21 case to $\sim 1.61 {\rm nK^2}$ in the SPT-3G case.

Incorporating the \planck{}'s $\tau$ dataset in the likelihood, error bars on $z_{50}$ and $\Delta z$ improve from 0.69 and 0.40 (Planck+R21) to 0.59 and 0.34 (Planck+SPT-3G) respectively. $\tau$ constraints from Planck+SPT-3G 
will correspond to the most sensitive inference on $\tau$ possible with current CMB experiments with error bars of $\sim 0.0052$. A comparison of the redshift evolution of ionizing source parameters and the evolution of ionized mass fraction has been presented in the right column in Figure \ref{fig:68pc}.
It's vital to recognize that the predictions made using the kSZ-only probe lean towards an earlier reionization scenario, contrasting with the scenarios favored by models that combine both the $\tau$ and kSZ probes. This leads to a marginally higher mean for the inferred $\tau$ compared to that inferred when both probes are utilized. Such an outcome is attributable to our chosen prior on reionization's end at $z = 5$ and on unphysical ionizing efficiencies. With the inclusion of $\tau$ constraint, models of reionization that start very early are rejected, limiting the kSZ contribution arising in the post-reionization era. Finally, we forecast errors on the patchy probes of reionization, $D^{\rm kSZ}_{\ell=3000} (\mu {\rm K^2})$ at $\sim 0.12$ and $D^{BB,{\rm reion}}_{\ell=200}({\rm nK^2})$ at $\sim 1.33$. The ability to place stringent constraints on the patchiness in the reionization era and hence the corresponding patchy $B$-mode polarization has important ramifications when examining the bias in detecting primordial B-mode signal \citep{2019MNRAS.486.2042M,jain23,2024MNRAS.527.2560J}. This furthers the need for detecting the kSZ signal over a broad multipole span through the Cross-ILC technique.

\subsection{Forecasts for kSZ extracted using Cross-ILC technique for
upcoming telescopes}\label{subsec:forecastcilc}
Upcoming CMB experiments will enable high-fidelity measurements of CMB probes of reionization. LiteBIRD (expected launch: $\sim$ 2028) and PICO (expected launch: sometime in next decade) aim to make E-mode polarization observations at the reionization bump ($\ell<20$),  to constraint $\tau$ at $\sigma_\tau=0.002$. Meanwhile, the best kSZ-measurements are anticipated from ground-based observatories, specifically the Simons Observatory (expected first-light: $\sim 2024$) and CMB-S4 (expected first-light: $\sim 2030$). The associated Cross-ILC error bars for these experiments have been presented in Section \ref{sec:compare_const_spt}.  Considering the upcoming
mission timelines, we propose the following combination of mock
data sets to forecast constraints on the reionization model

\begin{itemize}
    \item Planck+SO-Goal: Measurement of $\tau$ with \planck{} and fiducial measurement of kSZ corresponding to the goal configuration of Simons Observatory
    \item LB+S4-Wide:  fiducial measurement of $\tau$ with LiteBIRD and fiducial kSZ-power spectrum with CMB-S4 for their Wide configuration.
\end{itemize}
Further to emphasize the forecasts entailed by the kSZ extracted using the Cross-ILC technique we present the cases SO-Goal and S4-Wide. This corresponds to the forecast on our reionization model using just the upcoming kSZ fiducial data.

We present a comparison of forecasts on our model of reionization in Table \ref{tab:forecast-CROSS_ilc_COMPARISON_upcoming}. The two-dimensional posterior distribution for the cases SO-Goal and S4-Wide is presented in  Figure \ref{fig:alljustkszfig4a} and for the case Planck+SO-Goal and LB+S4-wide have been presented in Figure \ref{fig:alltaukszfig4b}. The Cross-ILC error bars obtained for Simons Observatory and SPT-3G are fairly consistent (refer Figure \ref{fig:jvarkSZ}) and hence, the error bars obtained on our reionization parameters are also roughly consistent. S4-Wide will enable the most sensitive small-scale temperature measurements enabling the most stringent constraints on our reionization model allowed by the CMB small-scale measurements. With S4-Wide measurements, $\tau$ is forecasted with error bars of $\sim 0.0056$, and the duration of reionization is constrained with error bars of $\sim 0.35$. Tight constraints on ionizing source properties will enable stringent constraints on the allowed patchiness during reionization and will enable constraints on $D^{BB}_{\ell=200} ({\rm nK^2})$ at $\sim 1.18$. 
  \setlength{\tabcolsep}{0.8pt}
\begin{table}
    \fontsize{7.0}{10}\selectfont
    \centering
    \caption{Comparison of forecasts (68\% limits) for reionization model parameters obtained from MCMC analysis using improved Cross-ILC error bars for SO-Goal and S4-Wide has been presented. In this regard, we present the following case: Planck+SO-Goal, and the LB+CMB-S4 and to emphasize the improvement entailed by the Cross-ILC errors we present the cases for SO-Goal and S4-Wide. The first four rows correspond to the free parameters of the model while the rest of the parameters are the derived parameters. The second column refers to the model of reionization used to generate the mock data for forecasts.}
    %|p{2.2cm}|p{0.6cm}|p{2.4cm}|p{2.6cm}|
    \begin{tabular}{|c|c|c|c|c|c|c|c|c|}
    \hline
    Data  &  & SO-goal & S4-Wide & Planck+SO-Goal & LB+S4-Wide \\
    Parameter  &Input & $68\% $ limits    &  $68\% $ limits&  $68\% $ limits&  $68\% $ limits \\
    \hline
    $\log_{10} \zeta_0$& 1.58 &    $2.03^{+0.56}_{-0.91}$ & \pc{2.09}{0.50}{0.85}&\pc{1.86}{0.48}{0.81} &\pc{1.73}{0.69}{0.29}   \\[0.075cm]
    $\log_{10} M_{\mathrm{min,0}}$ & 9.73 & $9.78^{+1.00}_{-0.37}$  &       \pc{9.86}{0.76}{0.28}&\pc{9.77}{0.99}{0.38}& \pc{9.81}{0.69}{0.29}\\[0.1cm]
    $\alpha_\zeta$ & -2.01 &  $-3.83^{+2.59}_{-2.36}$  & \pc{-4.16}{2.55}{2.09}&\pc{-4.40}{2.62}{2.37} & \pc{-3.63}{1.64}{1.42}\\[0.1cm]
    $\alpha_M$ & -2.06 &  $>-2.33$ &$>-2.38$ &$>-2.58$& $>-3.57$  \\[0.1cm]
    \hline
    Derived  &&\\
    Parameters&&\\
    $\tau$ & 0.054 &  $0.0593^{+0.0071}_{-0.0053}$&\pc{0.0602}{0.0062}{0.0050} & \pc{0.0567}{0.0054}{0.0048}& \pc{0.0550}{0.002}{0.002} \\[0.1cm]
    $z_{\mathrm{50}}$& 7.27 &  \pc{7.87}{0.84}{0.53}  &\pc{7.91}{0.74}{0.55}& \pc{7.61}{0.64}{0.52}& \pc{7.41}{0.26}{0.23} \\[0.1cm]
    $\Delta z$    & 1.31 & $1.11^{+0.24}_{-0.49}$  &\pc{1.07}{0.22}{0.48} & \pc{1.10}{0.21}{0.40} & \pc{1.20}{0.13}{0.29} \\[0.1cm]
    $D^{\mathrm{kSZ}}_{l=3000}$($\mu K^2$) & 3.00 &  $3.02^{+0.14}_{-0.15}$ & \pc{3.02}{0.04}{0.03}& \pc{2.98}{0.14}{0.13} & \pc{3.01}{0.03}{0.03}\\[0.1cm]
    $D^{\mathrm{kSZ,reion}}_{l=3000}$($\mu K^2$) & 1.33 & \pc{1.14}{0.22}{0.33}&$1.11^{+0.21}_{-0.30}$& \pc{1.19}{0.18}{0.30} & \pc{1.25}{0.09}{0.12}\\[0.1cm]
    $D^{\mathrm{kSZ,post-reion}}_{l=3000}$($\mu K^2$) & 1.67 & \pc{1.88}{0.35}{0.30}&$1.91^{+0.30}_{-0.25}$& \pc{1.79}{0.27}{0.21} & \pc{1.75}{0.14}{0.10}\\[0.1cm]
    $D^{BB,\mathrm{reion}}_{\ell=200}$(${\rm n K^2}$)   & 7.03 &  $7.07^{+0.93}_{-2.13}$  & \pc{7.07}{1.04}{1.32} & \pc{6.24}{0.95}{1.74}& \pc{6.59}{0.78}{1.02}\\[0.1cm]
    \hline
    \end{tabular}
    \label{tab:forecast-CROSS_ilc_COMPARISON_upcoming}
\end{table}
Combining  {\it Planck}'s measurement of $\tau$ for the case Planck+SO-Goal, unsurprisingly, results in error bars on our reionization model's free and derived parameters consistent with the forecast for Planck+SPT-3G. With the inclusion of $\tau$ measurement with LiteBIRD,
 the data set LB+S4-Wide will enable unprecedented constraints on our reionization model, representing the most stringent constraints on a reionization model with the combination of CMB polarization and temperature data. We observe that the error bars on the optical depth will be governed by  LiteBIRD's sensitivity and small-scale temperature information may not have further constraining power. We find that the mean redshift of reionization will be constrained with error bars of $\sim 0.25$ and the duration of reionization with an error bar of $\sim 0.21$. The combination of information on reionization history through $\tau$ and the information on patchiness through kSZ will reduce error bars on $D^{BB, {\rm reion}}_{\ell=200} ({\rm nK^2})$ to $\sim 0.90$. 
 
  Finally, to motivate the need to have multiple-bin detection of kSZ power spectrum we compare forecasts for two bin sizes, $\Delta \ell=250$ and $\Delta \ell=2500$ for the case S4-Wide in Appendix \ref{sec:compare_const_binchange}. We find that with access to the shape of the power spectrum, multiple bin detection of the power spectrum will enable tighter constraints on $\log_{10} M_{min,0}$ compared to a single data point at $\ell=3000$. This is intuitive as the shape of the signal correlates with the minimum mass of halo-hosting ionizing sources. However, because of the choice of a smooth fiducial power spectrum, improvement in $\log_{10} M_{min,0}$ does not entail a substantial improvement in reionization history parameters. Nevertheless, this result highlights the sensitivity of kSZ to the morphology of the sources and the potential to make further detailed inferences when combined with other probes sensitive to morphology e.g. 21cm signal. The above finding ascertains that, in the future, kSZ will play a significant role in constraining the patchy-B mode signal, possibly enabling the first detection of the patchy B-mode signal through the combination of small-scale kSZ signal, large-scale $E$-mode polarization, and $B$-mode polarization data sets.
 \begin{figure}
	\includegraphics[width=\columnwidth,trim={1.1cm 1.3cm 1.7cm 0.6cm}]{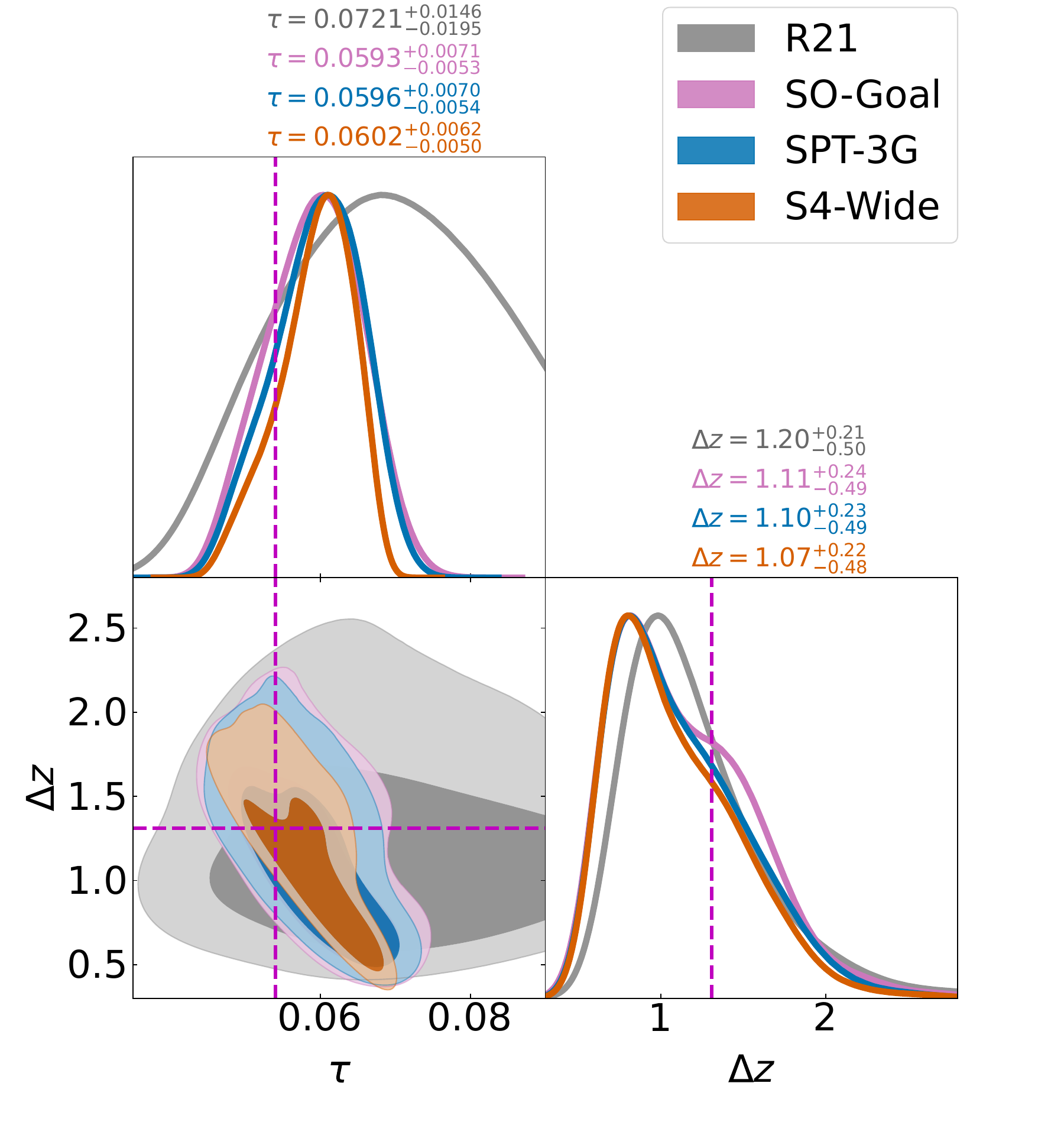}
    \caption{Comparison  of the 2D posterior distribution of optical depth $\tau$ and duration of reionization $\Delta z$ obtained from MCMC analysis using current observations from R21 and forecasts for SPT-3G, SO-Goal, and CMB S4-wide. The dashed lines denote the input model used to generate the fiducial data for forecasting.}
    \label{fig:taudeltazallksz}
\end{figure}
\begin{figure}
	\includegraphics[width=\columnwidth,trim={1.1cm 1.3cm 1.7cm 0.6cm}]{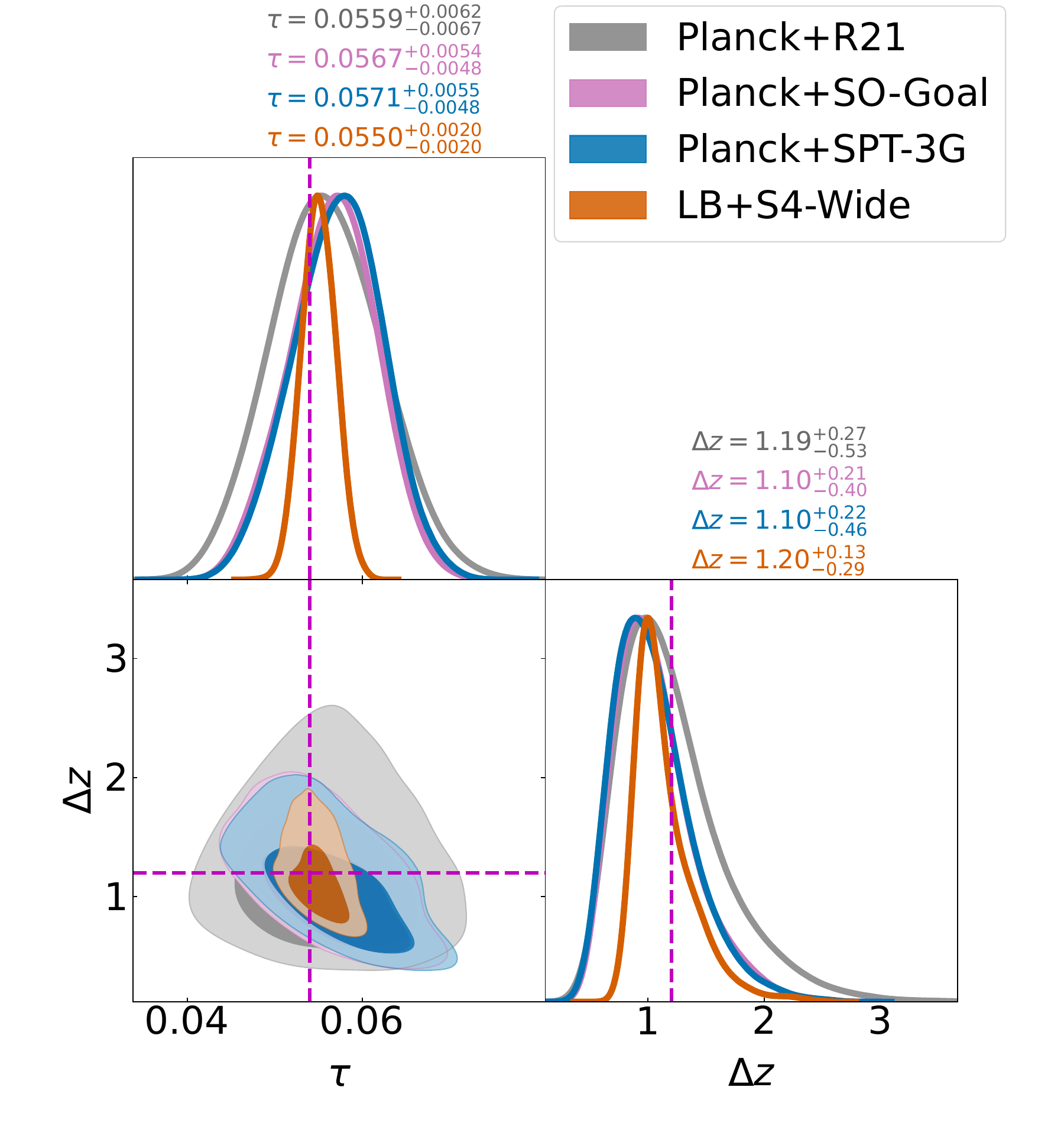}
    \caption{Comparison  of 2D posterior distribution of optical depth $\tau$ and duration of reionization $\Delta z$ obtained from MCMC analysis using current observations from Planck+R21 and forecasts for Planck+SPT-3G, Planck+SO-Goal, and LB+S4-wide. The dashed lines denote the input model used to generate the fiducial data for forecasting.}
    \label{fig:taudeltazalltauksz}
\end{figure}

\section{Discussion and Conclusion}\label{sec:discussandconclude}
Inhomogeneous scattering of CMB photons in the patchy reionization era modifies CMB properties. kSZ signal, a secondary temperature anisotropy, is one of the signature consequences of this interaction. The shape and amplitude of the kSZ signal encapsulate insights into the complex evolution of the reionization process. Current and Stage-4 CMB experiments intend to make available the most sensitive measurements of CMB temperature anisotropies enabling unprecedented detection of the kSZ component. However, the information gained will hinge on unbiased kSZ extraction. Standard template-based foreground removal techniques suffer from biases and in this regard, the novel Cross-ILC foreground removal enables access to the shape of the kSZ power spectrum.

In this work, we exploit the Cross-ILC extraction of kSZ, to forecast constraints on the reionization process. We show that our physically motivated model of reionization based on \script{}  has the potential to evaluate the complete shape of the kSZ power spectrum. We discuss the dependence of the shape of kSZ on various reionization parameters and find that measurement of the kSZ, through Cross-ILC, across multipole bins will enable potential insights into ionization history and the patchy topology of the reionization era. We confront our model of reionization with a simulated SPT-3G data set and find that even for the observed SPT-3G data, kSZ extracted via Cross-ILC would enable stringent constraints on our reionization model parameters without the need to invoke prior on $\tau$. The $\tau$ in such a case would be inferred with an error bar of $\sim 0.006$. This indicates the potential of kSZ data to provide a complementary means other than the large-scale CMB data to constrain the optical depth. The inclusion of available constraints on optical depth from \planck{} with Cross-ILC SPT-3G data will enable a constraint on the midpoint of reionization with an error bar at $\sim 0.59$ and on the duration of reionization with an error bar of $\sim 0.34$. These forecasts represent the best possible constraints on the reionization era that can be achieved, utilizing existing CMB data in conjunction with the extraction of kSZ through the Cross-ILC technique. With high-fidelity Stage-4 experiments, these constraints are shown to improve further, with the error bars on the reionization midpoint going as low as $\sim 0.25$ and the error on the duration of reionization improving to $\sim 0.21$. We summarize the potential of our model to capitalize on the recovery of kSZ through Cross-ILC in Figure \ref{fig:taudeltazallksz} where we present the 2D posterior distribution for the parameters $\tau$ and $\Delta z$ corresponding to the kSZ forecast cases we discuss in our work. We find that while improving generations of experiments targeting kSZ improves our ability to constrain the optical depth $\tau$, the constraints on the duration of reionization improve marginally. This significantly improves when we include the $\tau$ probe in the likelihood, where $\tau$ measurement improves our ability to constrain the reionization history, evident through tighter constraints on $\Delta z$, captured in Figure \ref{fig:taudeltazalltauksz}. 

Further, access to the shape of the kSZ power spectrum through Cross-ILC recovery would enable unprecedented constraints on patchy reionization signal. The most stringent forecast on error bar of $D^{\rm kSZ,reion}_{\ell=3000}$ and $D^{BB,{\rm reion}}_{\ell=200}$ being at $\sim 0.11 {\rm \mu K^2}$ and $\sim 0.9 {\rm nK^2}$ , when combining future $\tau$ measurements with LiteBIRD and kSZ power spectrum measurements with CMB-S4. This high-fidelity kSZ recovery could enable the first evidence of the patchy B-mode signal in the low-$\ell$ $B$-mode data through joint constraints  using the combination of small-scale kSZ
signal, large-scale $E$-mode polarization, and $B$-mode polarization
data sets. This has important ramifications in the unbiased detection of primordial $B$-mode signal \citep{2019MNRAS.486.2042M,jain23}.

In summary, this is the first work to forecast on the astrophysics of reionization using simulated kSZ in the range $\ell \in [2500,5000]$ with realistic error bars for a physically motivated model.
Throughout our analysis, we find that the improved S/N and access to the shape of kSZ enable unprecedented constraints on the patchy properties of reionization. These constraints will be complementary
to constraints from other upcoming probes of reionization, e.g., the 21 cm signal \citep{2015aska.confE..10M,2016JApA...37...29C}. Joint constraints \citep{2022PhRvD.105h3503B} as well as cross-correlation studies of kSZ with redshift-based 21-cm data \citep{2018MNRAS.476.4025M,2020ApJ...899...40L,2020arXiv201209851H} from upcoming telescopes will enable useful avenues to study the ionization topology as well as large-scale properties at reionization redshifts.

%%%%%%%%%%%%%%%%%%%%%%%%%%%%%%%%%%%%%%%%%%%%%%%%%%

\section*{Acknowledgments}
DJ and TRC acknowledge the support of the Department of Atomic
Energy, Government of India, under project no. 12-R\&D-TFR-
5.02-070. The work of SM is a part of the $\langle \texttt{data|theory}\rangle$ \texttt{Universe-Lab} which is supported by the TIFR and the Department of Atomic Energy, Government of India. 
SR acknowledges support from the Center for AstroPhysical Surveys (CAPS) at the National Center for Supercomputing Applications (NCSA), University of Illinois Urbana-Champaign. 

This work made use of the following computing resources: Illinois Campus Cluster, a computing resource that is operated by the Illinois Campus Cluster Program (ICCP) in conjunction with the National Center for Supercomputing Applications (NCSA) and which is supported by funds from the University of Illinois at Urbana-Champaign.

 \section*{Data Availability}
A basic version of the semi-numerical code SCRIPT for generating the
ionization maps and computing kSZ power spectrum, as used in the paper is publicly available at \url{https://bitbucket.org/rctirthankar/script}. Additionally, the bandpower errors from cross-ILC kSZ extraction corresponding to different CMB experiments, employed for Bayesian forecasting in this work, are publicly available at \url{https://github.com/sriniraghunathan/cross_ilc_methods_paper}.
Any other data related to the paper will be shared on reasonable request to the corresponding author (DJ).

\bibliographystyle{mnras}
\bibliography{examplev2}
\appendix
\section{Reionization Forecasts Without Gaussian Random Noise in Mock data}
\setlength{\tabcolsep}{8pt}
\begin{table*}
    %\fontsize{10}{10}\selectfont
    \textnormal
    \centering
    \caption{Comparison of forecasts (68\% limits) for reionization model parameters obtained from MCMC analysis from improved Cross-ILC error bars for SPT-3G, S0-goal and S4-wide for the NO GRN case. The No GRN refers to the case where Gaussian random realization of error was not added to the input kSZ data. The first four rows correspond to the free parameters of the model while the rest of the parameters are the derived parameters. The second column refers to the model of reionization used to generate the mock data. }
    %|p{2.2cm}|p{0.6cm}|p{2.4cm}|p{2.6cm}|
    \begin{tabular}{|c|c|c|c|c|c|c|c|c|}
    \hline
    Parameter  &  Input & SPT-3G  & SO-Goal & S4-Wide& Planck+SPT-3G & Planck+SO-Goal & LB+S4-wide \\
    \hline\hline
    \multicolumn{8}{l}{{\it Baseline parameters}} \\
    \hline
    $\log_{10} \zeta_0$& 1.58 & \pc{1.94}{0.46}{0.86} & \pc{1.94}{0.48}{0.91}  & \pc{1.70}{0.33}{0.61}& \pc{1.77}{0.38}{0.75}   & \pc{1.75}{0.38}{0.75}  & \pc{1.50}{0.39}{0.36}  \\[0.075cm]
    $\log_{10} M_{\mathrm{min,0}}$ & 9.73  & \pc{9.66}{1.03}{0.45} & \pc{9.63}{1.06}{0.43}  & \pc{9.43}{0.85}{0.43}& \pc{9.74}{0.97}{0.38}   &    \pc{9.64}{0.95}{0.43}  & \pc{9.50}{0.75}{0.34}\\[0.1cm]
    $\alpha_\zeta$ & -2.01 & \pc{-3.93}{2.43}{2.47}  & \pc{-4.02}{2.58}{2.14}  & \pc{-3.62}{1.90}{2.24}& \pc{-4.42}{2.57}{2.13}  & \pc{-4.41}{2.55}{2.06} & \pc{-3.52}{1.68}{1.21}\\[0.1cm]
    $\alpha_M$ & -2.06  & $>-2.37$  & $>-2.41$ & $>-2.80$    & $>-2.89$   &     $>-2.92$ & $>-3.60$   \\[0.1cm]
    \hline\hline
    \multicolumn{8}{l}{{\it Derived parameters}} \\
    \hline    
    $\tau$ & 0.054  & \pc{0.0595}{0.0070}{0.0048} & \pc{0.0597}{0.0067}{0.0056}& \pc{0.0584}{0.0053}{0.0052}& \pc{0.0570}{0.0054}{0.0046}  & \pc{0.0568}{0.0055}{0.0050} & \pc{0.0549}{0.0019}{0.0020}  \\[0.1cm]
    $z_{\mathrm{50}}$& 7.27 & \pc{7.89}{0.84}{0.50}  & \pc{7.94}{0.79}{0.59} & \pc{7.59}{0.70}{0.43}& \pc{7.62}{0.65}{0.49} & \pc{7.60}{0.66}{0.53}& \pc{7.36}{0.25}{0.25} \\[0.1cm]
    $\Delta z$    & 1.31 & \pc{1.15}{0.20}{0.48}  & \pc{1.15}{0.22}{0.54}  & \pc{1.26}{0.30}{0.50}  & \pc{1.10}{0.23}{0.43}  & \pc{1.10}{0.22}{0.42}  & \pc{1.26}{0.13}{0.36}  \\[0.1cm]
    $D^{\mathrm{kSZ}}_{l=3000}$($\mu K^2$) & 3.00 & \pc{3.03}{0.13}{0.13} & \pc{3.04}{0.13}{0.14}  & \pc{3.01}{0.03}{0.03}& \pc{3.00}{0.12}{0.11}  &\pc{3.00}{0.13}{0.14} & \pc{3.00}{0.03}{0.03} \\[0.1cm]
    $D^{BB,\mathrm{reion}}_{\ell=200}$(${\rm n K^2}$)   & 7.03 & \pc{6.82}{0.95}{1.91} & \pc{6.78}{1.04}{1.88}  & \pc{6.49}{1.06}{1.51}& \pc{6.08}{1.02}{1.49}  & \pc{6.06}{1.02}{1.53} & \pc{6.41}{0.69}{0.87} \\[0.1cm]
    \hline
    \end{tabular}
    \label{tab:forecast-CROSS_ilc_COMPARISON_no_noise}
\end{table*}

In Table \ref{tab:forecast-CROSS_ilc_COMPARISON_no_noise} we present the forecasts on reionization parameters for the cases as presented in Section \ref{sec:compare_const}. The key deviation in this analysis is the exclusion of Gaussian random noise realization from the mock data set. Hence, these forecasts represent a hypothetical perspective that does not capture a realistic observing scenario.

\section{Posterior distributions from forecasts for kSZ extracted using Cross-ILC technique for  upcoming telescopes }\label{sec:compare_const_upcoming}

\begin{figure*}
    \centering
    \includegraphics[width=\linewidth]{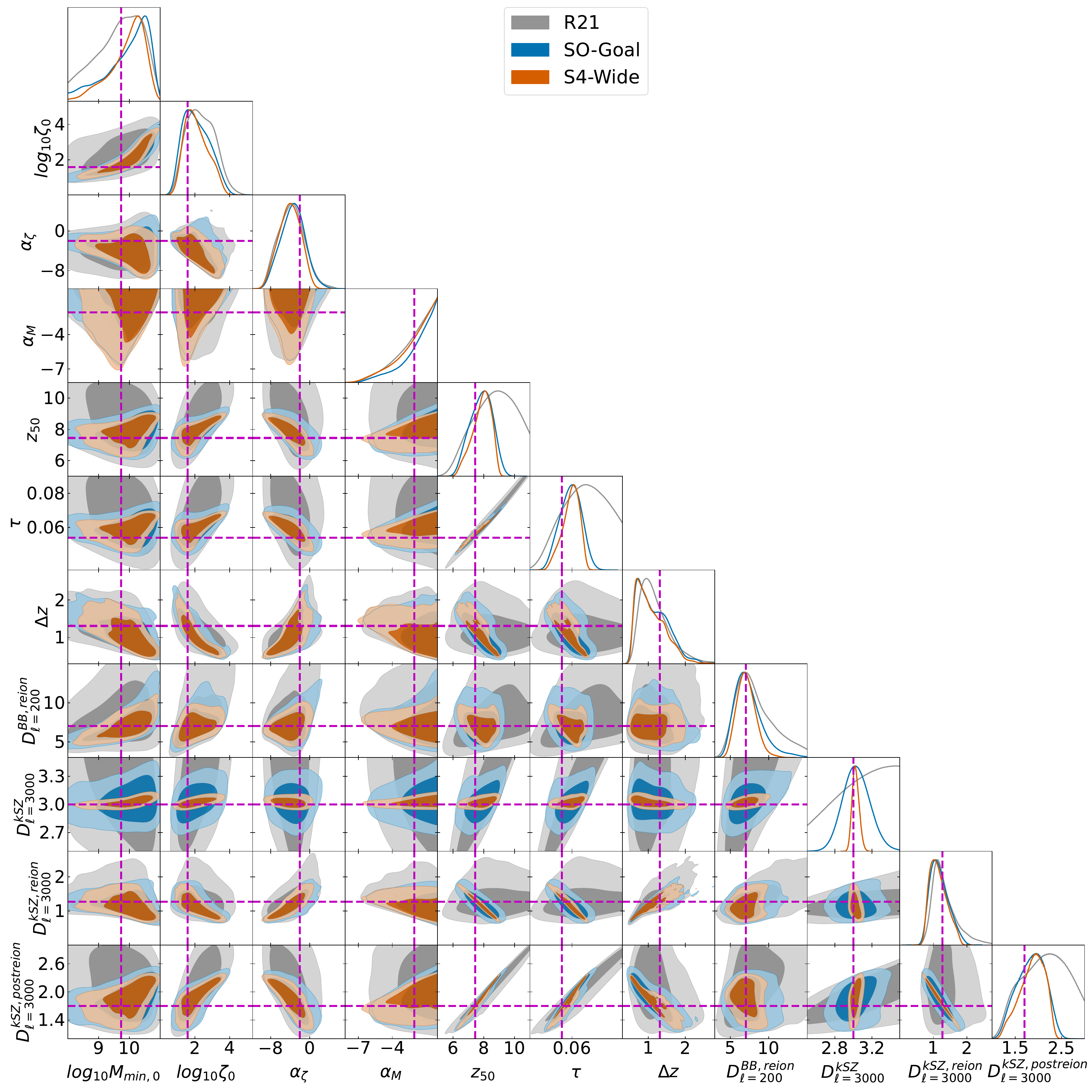}
    \caption{Comparison of the 2D posterior distribution of reionization model parameters obtained from MCMC analysis for the case SO-Goal, and S4-wide. For comparison, the 2D posterior distribution of reionization model parameters for the case R21 has been presented in grey. The dashed lines denote the input model used to generate the fiducial data for forecasting.}
    \label{fig:alljustkszfig4a}
\end{figure*}

\begin{figure*}
    \centering
    \includegraphics[width=\linewidth]{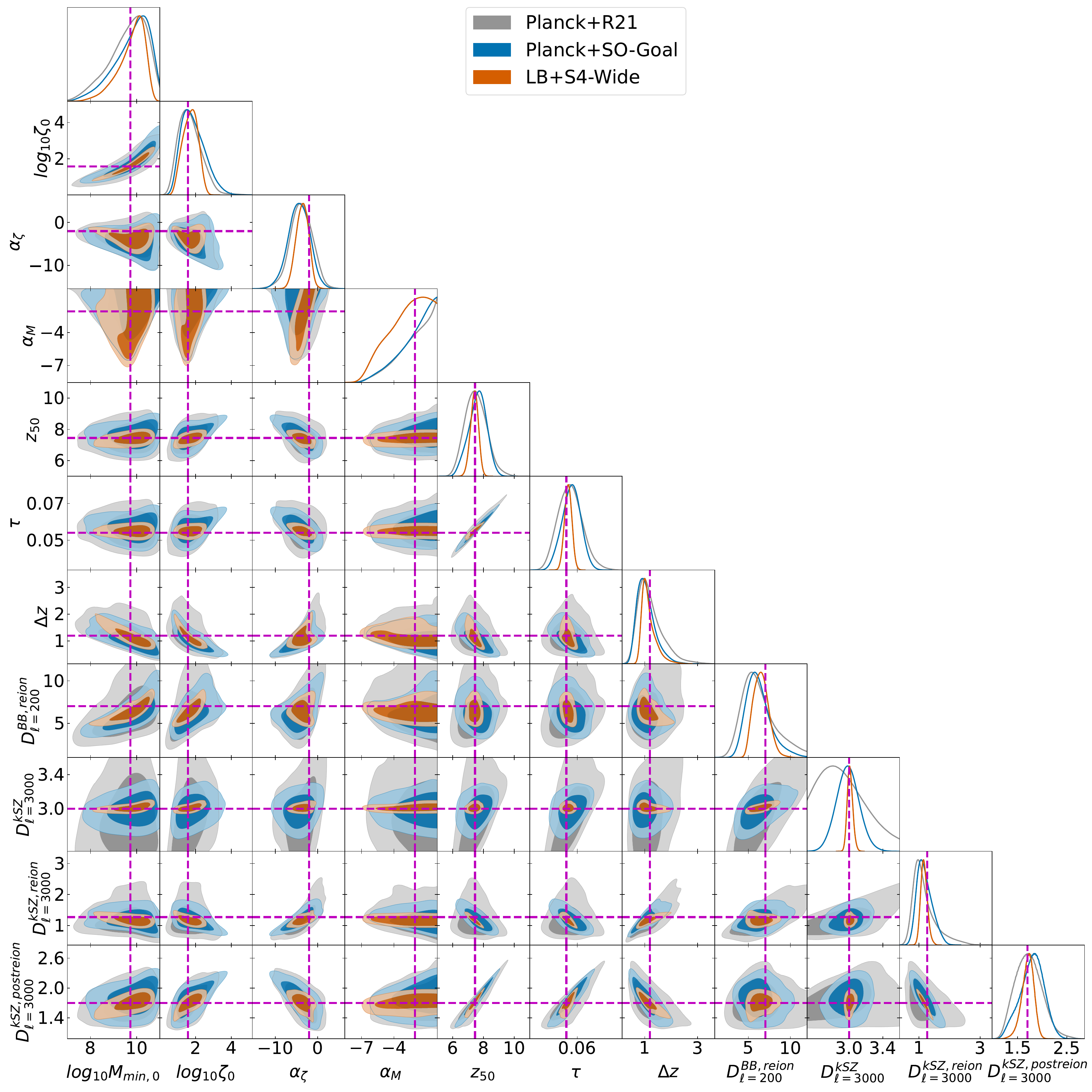}
    \caption{Comparison of 2D posterior distribution of reionization model parameters obtained from MCMC analysis the case Planck+SO-Goal
and forecasts for case LB+S4-Wide. For comparison, the 2D posterior distribution of reionization model parameters for the case Planck+R21 has been presented in grey. The dashed lines denote the input model used to generate the fiducial data for forecasting.}
    \label{fig:alltaukszfig4b}
\end{figure*}

In this section, we present the posterior distribution of reionization model parameters, derived from Bayesian forecasts using the anticipated kSZ and $\tau$ measurements. These measurements are expected from the high-fidelity temperature and polarization observations from the upcoming CMB experiments. A detailed discussion of these forecasts has been presented in Section \ref{subsec:forecastcilc}. The two-dimensional posterior distributions for the anticipated kSZ measurements from SO-Goal and S4-Wide are shown in Figure \ref{fig:alljustkszfig4a} while the forecasts from the joint data set combining kSZ and $\tau$ data through cases Planck+SO-Goal and LB+S4-Wide are depicted in Figure \ref{fig:alltaukszfig4b}.

\section{Parameter forecast comparison of single and multiple bin detection of kSZ extracted using Cross-ILC}\label{sec:compare_const_binchange}

\begin{figure*}
    \centering
    \includegraphics[width=\linewidth, trim={4cm 4cm 4cm 0cm}, clip]{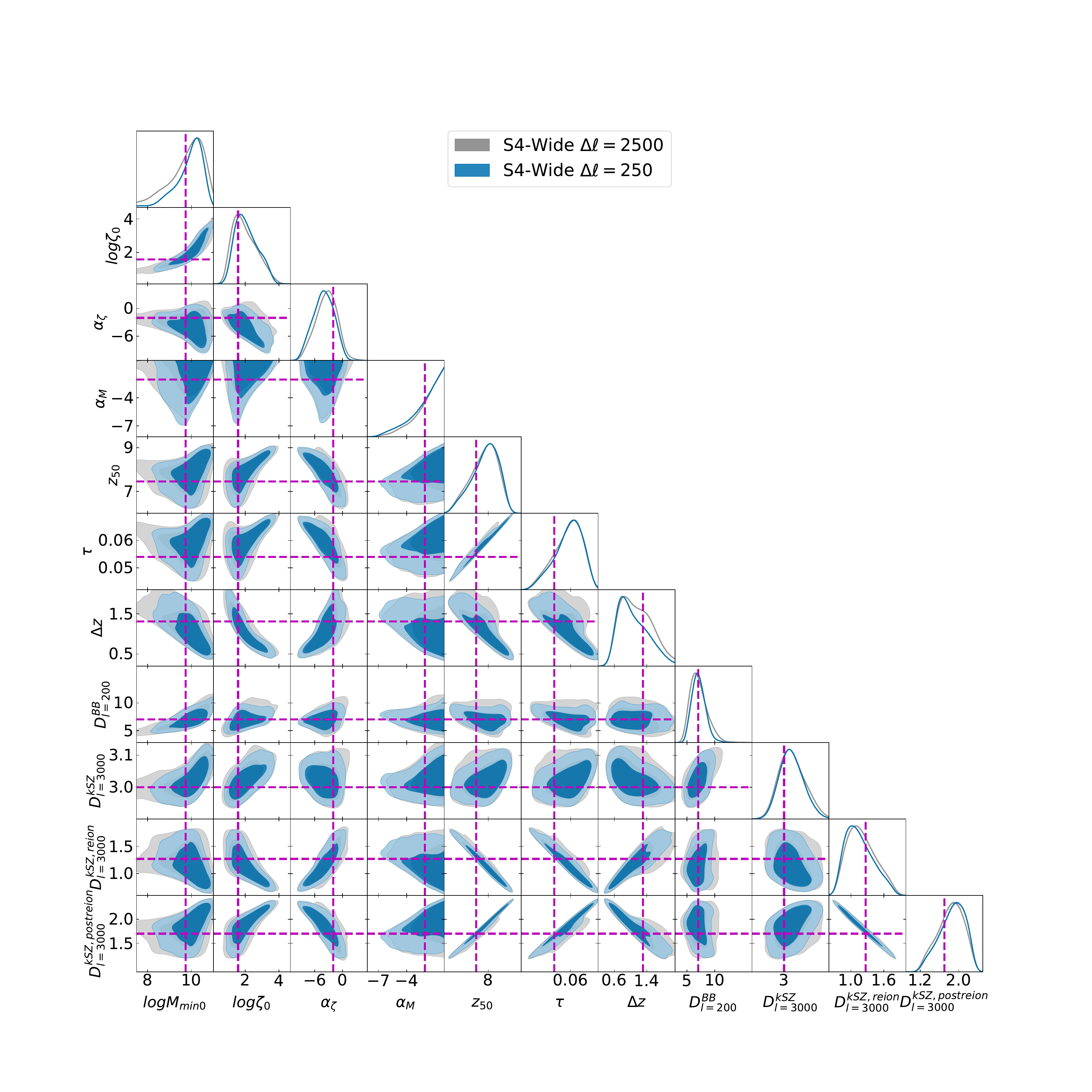}
    \caption{Comparison of the 2D posterior distribution of reionization model parameters obtained from MCMC analysis using forecasted kSZ observations for the case S4-Wide
for different bin width choices of $\Delta \ell=2500$ and $\Delta \ell=250$. The dashed lines denote the input model used to generate the fiducial data for forecasting.}
    \label{fig:alljustksz_s4wide_appendix_binchange}
\end{figure*}

\begin{figure}
    \centering
    \includegraphics[width=\columnwidth]{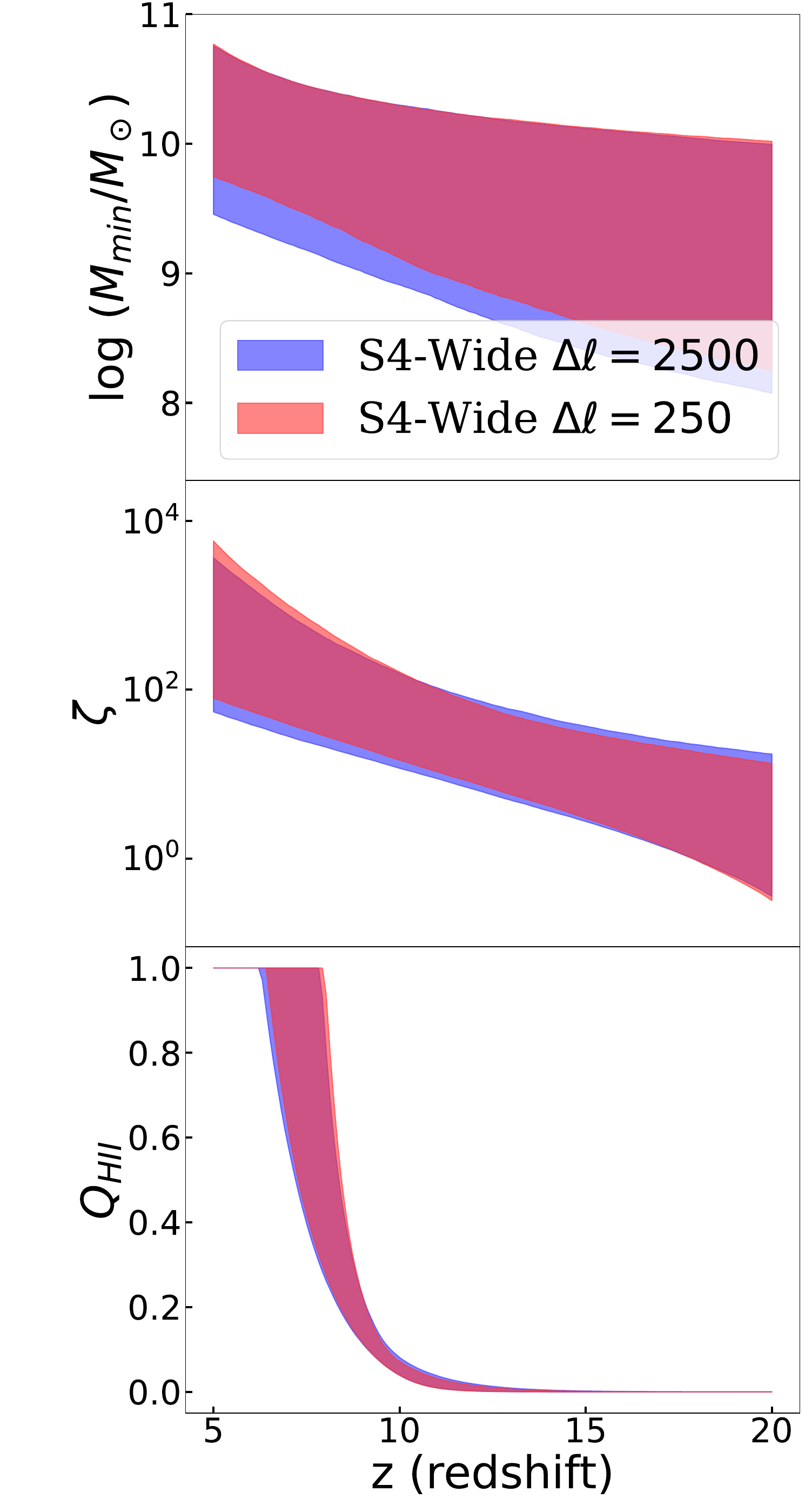}
    \caption{Evolution of minimum mass of haloes hosting sources which contribute to reionization, the ionizing efficiency of the sources $\zeta(z)$ and the ionized mass fraction $Q_{\rm HII}$ for models in MCMC chains corresponding to the case S4-Wide for different bin width detection $\Delta \ell=[250,2500]$.}
    \label{fig:enter-label}
\end{figure}

Compared to the stadard template fitting approaches of estimating the kSZ power spectrum at $\ell = 3000$ (\citetalias{reichardt21}, \citealt{gorce22}), which are sensitive to the assumptions about foreground signals, the Cross-ILC technique allows for a robust estimate of the kSZ \citepalias{raghunathan23}. Subsequently, it allows us to capture the kSZ power spectrum over a range of multipoles, which provides useful insights about the process of reionization, as motivated already in the main text in Sections \ref{sec:motivation} and \ref{sec:compare_const}. Here, we present a quantitative comparison of the single bin ($\Delta \ell = 2500$) vs the multiple bin ($\Delta \ell = 250$) detection of kSZ power spectrum. We limit this comparison to the S4-Wide experiment which returns the highest kSZ SNR. We present the constraints in Table \ref{tab:alljustksz_s4wide_appendix_binchange} and the posterior distributions in Figure \ref{fig:alljustksz_s4wide_appendix_binchange}. For $\Delta \ell=250$, we find the constraints on $\log_{10} M_{min,0}$ improves by $\sim30\%$ compared to a single bin detection, as demonstrated in the top panel in Figure \ref{fig:enter-label}. This improvement is expected, as the shape of the patchy reionization kSZ contribution correlates with the minimum mass of haloes that can host the ionizing sources, as depicted in the top panel of Figure \ref{fig:varksz_patchy}. For other parameters, we do not find a significant changes in the constraints between single- vs multiple-bin measurements of the kSZ power spectrum. Due to the relatively smooth nature of our fiducial power spectrum within the multipole range of $\ell \in [2500,5000]$ (Figure~\ref{fig:jvarkSZ}), the improved constraints on $\sigma(\log_{10} M_{min,0})$ does not substantially improve our understanding of ionization history through parameters like $z_{\rm 50}$,$\tau$ and $\Delta z$ (as shown in Table \ref{tab:alljustksz_s4wide_appendix_binchange}). However, the improvement in the $\log_{10} M_{min,0}$, coming from the shape of the power spectrum, will improve our understanding of the topology of the ionizing field, and hence highlights the need to capture the kSZ power spectrum across multiple bins. This will be compounded when conducting joint constraints with observables sensitive to the topology of reionization, e.g. 21cm signal.
%%%%%%%%%%%%%%%%%%%%%%%%%%%%%%%%%%%%%%%%%%%%%%%%%%

 \setlength{\tabcolsep}{4pt}
\begin{table}
    \fontsize{8.0}{10}\selectfont
    \centering
    \caption{Comparison of forecasts (68\% limits) for reionization model parameters obtained from MCMC analysis from improved Cross-ILC error bars for S4-Wide. We compare the obtained constraints for different bin width choices of $\Delta \ell=250$ and $\Delta \ell=2500$. }
    % 
    %|p{2.2cm}|p{0.6cm}|p{2.4cm}|p{2.6cm}|
    \begin{tabular}{|c|c|c|c|}
    \hline
    Parameter  &  Input & S4-Wide $\Delta \ell=2500$ & S4-Wide $\Delta \ell=250$ \\
    \hline%\hline
    $\log_{10} \zeta_0$& 1.58& \pc{1.98}{0.51}{0.92} & \pc{2.09}{0.50}{0.85}\\[0.075cm]
    $\log_{10} M_{\mathrm{min,0}}$ & 9.73 & \pc{9.74}{1.02}{0.35} & \pc{9.86}{0.76}{0.28}  \\[0.1cm]
    $\alpha_\zeta$ & -2.01 & \pc{-3.48}{2.52}{2.08} & \pc{-4.16}{2.55}{2.09} \\[0.1cm]
    $\alpha_M$ & -2.06 & $>-2.22$ & $>-2.38$      \\[0.1cm]
    \hline%\hline  
    $\tau$ & 0.054 & \pc{0.0601}{0.0064}{0.0052} & \pc{0.0602}{0.0062}{0.0050}  \\[0.1cm]
    $z_{\mathrm{50}}$& 7.27 & \pc{7.87}{0.74}{0.56} & \pc{7.91}{0.74}{0.55}  \\[0.1cm]
    $\Delta z$    & 1.31 & \pc{1.17}{0.26}{0.54} & \pc{1.07}{0.22}{0.48}     \\[0.1cm]
    $D^{\mathrm{kSZ}}_{l=3000}$($\mu K^2$) & 3.00 & \pc{3.03}{0.03}{0.05} & \pc{3.02}{0.04}{0.03}\\[0.1cm]
    $D^{\mathrm{kSZ,reion}}_{l=3000}$($\mu K^2$) & 1.33 & \pc{1.20}{0.23}{0.32} & \pc{1.11}{0.21}{0.30}   \\[0.1cm]
    $D^{\mathrm{kSZ,post-reion}}_{l=3000}$($\mu K^2$) & 1.67 & \pc{1.83}{0.35}{0.25} & \pc{1.91}{0.30}{0.25}   \\[0.1cm]
    $D^{BB,\mathrm{reion}}_{\ell=200}$(${\rm n K^2}$)   & 7.03 & \pc{7.10}{1.09}{1.85} & \pc{7.07}{1.04}{1.32}    \\[0.1cm]
    \hline
    \end{tabular}
    \label{tab:alljustksz_s4wide_appendix_binchange}
\end{table}

% Don't change these lines
\bsp	% typesetting comment
\label{lastpage}
\end{document}